\documentclass[fleqn,10pt,twocolumn]{wlscirep}
\usepackage[utf8]{inputenc}
\usepackage{url}
\usepackage{hyperref}

\usepackage[T1]{fontenc}
\usepackage{bm,ulem}

\usepackage{verbatim,todonotes}
\usepackage{csquotes}
\usepackage{epsf,epsfig,amsmath,amssymb,dsfont}
\usepackage{amsthm,mathrsfs} 
\usepackage{graphicx} 
\usepackage{float}
\usepackage{float}
\usepackage{hyperref} 
\usepackage[]{subfigure}
\usepackage{multirow}
\usepackage{mathtools}
\usepackage{color}
\usepackage[framemethod=TikZ]{mdframed}
\mdfdefinestyle{MyFrame}{%
    linecolor=blue,
    outerlinewidth=2pt,
    roundcornehttps://www.overleaf.com/project/5f75a964ef1fd40001abe1d2r=20pt,
    innertopmargin=\baselineskip,
    innerbottommargin=\baselineskip,
    innerrightmargin=20pt,
    innerleftmargin=20pt,
    backgroundcolor=gray!50!white}
\usepackage[font=small,labelfont=bf,
   justification=justified,
   format=plain]{caption}
\usepackage{cleveref}

\DeclareMathOperator{\sign}{sign}

\newcommand{\T}[1]{\text{\tiny{#1}}}
\newcommand{\se}{\text{\tiny{$\mathcal{SE}$}}}
\newcommand{\s}{\text{\tiny{$\mathcal{S}$}}}
\newcommand{\e}{\text{\tiny{$\mathcal{E}$}}}
\newcommand{\I}{\text{\tiny{I}}}
\newcommand{\R}{\text{\tiny{R}}}
\newcommand{\RI}{\text{\tiny{R/I}}}
\newcommand{\be}{\begin{equation}}
\newcommand{\ee}{\end{equation}}
\newcommand{\baln}{\begin{align}}
\newcommand{\ealn}{\end{align}}
\newcommand{\ben}{\begin{equation*}}
\newcommand{\een}{\end{equation*}}

\newcommand{\MP}[1]{\textcolor{black}{#1}}

\long\def\symbolfootnote[#1]#2{\begingroup%
\def\thefootnote{\fnsymbol{footnote}}\footnote[#1]{#2}\endgroup}

\newcommand{\ket}[1]{\left| {#1} \right\rangle}
\newcommand{\bra}[1]{\left\langle {#1}\right|}
\newcommand{\braket}[1]{\langle {#1} \rangle}

\title{\Large{Testing the foundations of quantum physics in space}\\ 
\large{Interferometric and non-interferometric tests with Large Particles}}

\author[1,2,*,+]{Giulio Gasbarri}

\author[3,4,*,+]{Alessio Belenchia}

\author[5,6]{Matteo Carlesso}

\author[6,7]{Sandro Donadi}

\author[5,6]{Angelo Bassi}
\author[8,9]{Rainer Kaltenbaek}
\author[4]{Mauro Paternostro}
\author[2]{Hendrik Ulbricht}

\affil[1]{F\'isica Te\`orica: Informaci\'o i Fen\`omens Qu\`antics, Department de F\'isica, Universitat Aut\`onoma de Barcelona, 08193 Bellaterra (Barcelona), Spain}
\affil[2]{School of Physics and Astronomy, University of Southampton,Southampton SO17 1BJ, United Kingdom}
\affil[3]{Institut f\"{u}r Theoretische Physik, Eberhard-Karls-Universit\"{a}t T\"{u}bingen, 72076 T\"{u}bingen, Germany}
\affil[4]{Centre for Theoretical Atomic, Molecular, and Optical Physics, School of Mathematics and Physics, Queens University, Belfast BT7 1NN, United Kingdom}
\affil[5]{Department  of  Physics,University  of  Trieste,Strada  Costiera  11,  34151  Trieste,Italy}
\affil[6]{Istituto  Nazionale  di  Fisica  Nucleare,Trieste  Section,  Via  Valerio  2,  34127  Trieste,Italy} 
\affil[7]{Frankfurt Institute for Advanced Studies (FIAS),Ruth-Moufang-Stra{\ss}e 1, 60438 Frankfurt am Main, Germany}
\affil[8]{Faculty  of  Mathematics  and  Physics,University  of  Ljubljana,Jadranska  ulica  19,  1000  Ljubljana,Slovenia}
\affil[9]{Institute  for  Quantum  Optics  and  Quantum  Information,  Vienna,Austria}

\affil[*]{corresponding autors: giulio.gasbarri@uab.cat,alessio.belenchia@uni-tuebingen.de}
\affil[+]{these authors contributed equally to this work}

\begin{abstract}
Quantum technologies are opening novel avenues for applied and fundamental science at an impressive pace. In this perspective article, we focus on the promises coming from the combination of quantum technologies and space science to test the very foundations of quantum physics and, possibly, new physics. In particular, we survey the field of mesoscopic superpositions of nanoparticles and the potential of interferometric and non-interferometric experiments in space for the investigation of the superposition principle of quantum mechanics and the quantum-to-classical transition. We delve into the possibilities offered by the state-of-the-art of nanoparticle physics projected in the space environment and discuss the numerous challenges, and the corresponding potential advancements, that the space environment presents. In doing this, we also offer an ab-initio estimate of the potential of space-based interferometry with some of the largest systems ever considered and show that there is room for tests of quantum mechanics at an unprecedented level of detail. 
\end{abstract}

\begin{document}

\flushbottom
\maketitle

\thispagestyle{empty}


\section{Introduction}\label{introduction}
Quantum mechanics is one of the most successful physical theories human kind has ever formulated. Nonetheless, its interpretation and range of validity eludes our full grasping. One of the basic features of quantum physics is the superposition principle which, when applied to the macroscopic world, leads to counter-intuitive states akin to the celebrated Sch\"odinger's cat. While 
models beyond quantum mechanics, challenging some of its interpretational issues, have been formulated in its early days, testing the predictions of the theory when applied to the macroscopic world has proven to be a tall order. The main reason for this is the intrinsic difficulty in isolating large systems from their environment. 

Space offers a potentially attractive arena for such an endeavour, promising the possibility to create and verify the quantum properties of macroscopic superpositions far beyond current Earth-based capabilities~\cite{kaltenbaek2012macroscopic,kaltenbaek2016macroscopic,2018cosp...42E1659K,QPPF}. In this work, we focus on the efforts to test the boundaries of quantum physics in space employing nanoparticles, which are one of the best suited candidate for quantum superpositions of high-mass objects. \MP{It should be noticed that, while we will focus on testing quantum physics, large spatial superpositions of massive systems are bound to be \MP{sensitive probes for many other} physical phenomena, from dark matter and dark energy searches~\cite{riedel2013direct, bateman2015existence, riedel2017decoherence, carney2019ultralight, carney2020mechanical, monteiro2020search,khoury2004chameleon, rider2016search, moore2014search} to gravimetry and Earth observation applications~\cite{qvarfort2018gravimetry,hebestreit2018sensing}.}

We delve into the possibilities offered by the state-of-the-art of nanoparticle physics projected in the space environment. In doing so, we offer an ab-initio estimate of the potential of space-based interferometry with some of the largest systems ever considered and show that there is room for testing quantum mechanics at an unprecedented level of detail. 

\MP{The remainder of this paper} is organized as follow. In Sec.~\ref{Ch1}, after a brief introduction to the problem at hand and its relevance in fundamental physics, we discuss the advantages potentially offered by a space environment for quantum experiments \MP{based on}  large quantum superpositions of nanoparticles. We also give a self-contained overview of the current state-of-the-art for space missions proposals. In Sec.~\ref{Applications}, we discuss a first class of experiments that can be performed in space, i.e., non-interferometric ones which do not require the creation of macroscopic superpositions and exploit the free-evolution spread of the position of a quantum particle. In Sec.~\ref{ProofOfPrincipleImlementation}, we consider interferometric experiments which, in contrast, require the  creation and verification of large superpositions but also offer the benefit of a direct test of both the superposition principle of quantum mechanics and of competing theories. In particular, we present here an ab-initio estimate of the potential of space-based interferometry with large nanoparticles. We conclude in Sec.~\ref{WishList} with a discussion of our results and an outlook to the evolution of the field.


\section{Superposition of macroscopic systems: the case for space}\label{Ch1}
The predictions of quantum physics have been confirmed with a high degree of precision in a multitude of experiments, from the sub-atomic scale up to matter-wave interferometry with tests masses of nearly $10^5$ atomic mass units (amu)~\cite{fein2019quantum}. The basis for observing matter-wave interference is the quantum superposition principle, one of the pillars of quantum physics. While quantum physics does not pose any fundamental limitation to the size of quantum superposition states, the Gedankenexperiment of the  Schr\"{o}dinger's cat~\cite{Schroedinger1935b} illustrates the controversies entailed by the superposition principle when extended to the macroscopic world. Many proposals have been formulated in an attempt to establish a mechanism that would lead to the emergence of a classical world at macroscopic scales. Among them we find Bohmian mechanics~\cite{bohm1952suggested,durr2009bohmian}, decoherence histories~\cite{griffiths1984consistent}, the many-world interpretation~\cite{everett1957relative} and collapse models~\cite{bassi2003dynamical,RevModPhys.85.471} to name a few. The latter differs from the other proposals in the fact that they predict a phenomenology that deviates from the one of standard  quantum mechanics, albeit in a delicate fashion. In this sense, collapse models  represent an alternative construction to standard quantum theory, more than an alternative interpretation recovering all the predictions of the latter.
\MP{In light of the central role that they play in the experimental investigation of quantum macroscopicity~\cite{PhysRevLett.110.160403,PhysRevLett.119.100403}, in the following, we will focus on such models as benchmarks for precision tests of quantum mechanics.}

{
In 2010, a proposal for experimentally creating and verifying a state akin to the one of Schr\"{o}dinger's  cat based on the use of massive mechanical resonators was put forward within the context of the MAQRO proposal~\cite{kaltenbaek2012macroscopic}. The latter put forward the vision of harnessing the unique environment provided by space to test quantum physics in a dedicated, medium-sized space mission to be conducted within the framework of the "Cosmic Vision programme" run by the European Space Agency (ESA). The scope of the endeavour was to create a macroscopic superposition of motional states of a massive particle and probe its quantum coherence by allowing the wave functions of the components of such superposition to interfere, as in a double-slit experiment. The space-based environment would guarantee unprecedented levels of protection from environmental noises, as well as favourable working conditions for the engineering of the cat-like state~\cite{kaltenbaek2012macroscopic}.}

Near-field interferometry has later been identified as a viable route \MP{for} the achievement of the original goals of MAQRO~\cite{kaltenbaek2016macroscopic}, holding the promises for testing the superposition principle with particles of mass up to $10^{11}$~amu. This would be at least six orders of magnitude larger than the current record~\cite{fein2019quantum}\MP{. It would also far exceed the projected upper-bound to the masses that could be used in similar ground-based experiments. \MP{Such terrestrial upper-bounds are strongly limited by the achievable free-fall times on Earth}~\cite{gasbarri2020prospects} (cf. Sec.~\ref{advantage}).} 

{
The basic payload consists of optically trapped dielectric nanoparticles with a target mass range from $10^7$ to $10^{11}$~amu. The main scientific objectives are to perform both near-field interferometric and non-interferometric experiments. In both cases, high-vacuum and cryogenic temperatures are needed. The particles, after loading, are initially trapped in an optical cavity and their center-of-mass degree of freedom cooled down by a 1064~nm laser entering the quantum regime. For this purpose, two TEM$_{00}$ modes  with  orthogonal  polarization are to be used for trapping and side-band cooling along the cavity axis. The transverse motion is instead cooled employing a TEM$_{01}$ and a TEM$_{10}$ mode. After the initial state preparation, the particle can be released from the optical trap and undergoes different evolutions -- free fall expansion, coherent manipulations, and quantum detection -- depending on the experiments to be performed. 
}

{
The feasibility of the avenue identified in MAQRO has recently been investigated in a Quantum Physics Payload platForm (QPPF) study at the ESA Concurrent Design Facility~\cite{QPPF}. Such study has identified {\bf (a)} the core steps towards the realisation of a space-based platform for high-precision tests of quantum physics, and {\bf (b)} the potential of such platforms to test quantum physics with increasing test mass with the scope to ascertain potential deviations from the predictions of quantum physics due, for instance, to gravity. The ultimate goal of these endeavours is to provide a reference mission design for quantum-physics experiments in space.
}

{
The QPPF~\cite{QPPF} study culminated with the identification of a suitable combination of feasible free-fall times, temperature and pressures [cf. Table \ref{tab:parameterssmall}], setting a target of 2034 for the launch of the mission.  
}
{While several technical challenges remain to be addressed, the QPPF study has consolidated the intention to leverage on the expertise in near-field interferometry and optomechanics where state-of-the-art experiments with large molecule at near $10^5$~amu have been reported~\cite{fein2019quantum}, ground-state cooling achieved~\cite{delic2020cooling}, and proof-of-concept proposals for ground-based interferometry with large-mass experiments put forward~\cite{bateman2014near,gasbarri2020prospects}.  
In this respect, it should be mentioned that theoretical proposals based on other approaches, most noticeably magnetic levitation, have recently attracted the attention of part of the community~\cite{pino2018chip}. These proposals envision testing the superposition principle with ground-based experiments, overcoming some of the existing limitations through  
{the low noise level, long coherent-operation times, and lack of need for light fields driving the dynamics promised by magnetic levitation}. The use of masses of the order of $10^{13}$~amu has been forecast in this context~\cite{pino2018chip}. While extremely interesting, magnetic levitation technologies for quantum experiments are still at an early stage~\cite{rusconi2018levitated,
rusconi2017linear,rusconi2017quantum,druge2014damping} with, at present, no quantum superposition having been created which such techniques. }

{
\begin{table}[t]
    \centering
    \begin{tabular}{c|c}
    \hline
        \textbf{Parameters} & \textbf{Values}\\
        \hline
        \hline
         Free fall time $t$& up to $100\,$s  \\
         Environmental temperature $T_\text{env}$&down to 20\,K\\
         Pressure& down to $10^{-11}\,$Pa\\
         Masses& $10^{7} - 10^{11}$\,amu\\ 
         Diameters & $20 - 400$\,nm\\
    \hline
    \end{tabular}
    \caption{Possible combination of the parameters identified for the MAQRO mission and considered by ESA QPPF study~\cite{QPPF}.}
    \label{tab:parameterssmall}
\end{table}
}

\subsection{In search of the quantum-to-classical boundary}\label{subsec::QuantumClassical}

The extreme fragility of spatial quantum superpositions in the presence of environmental interactions (ubiquitous in any realistic setting) makes testing the superposition principle at macroscopic scales a tall order. Indeed, such interactions result in a suppression of quantum coherence in position that can be described by the following master equation in position representation\cite{joos1985emergence}
\begin{equation}\label{master_gen}
    \frac{d\langle x|\hat\rho_{t}|x'\rangle}{dt}=-\frac{i}{\hslash}\langle x|\left[\hat H,\hat\rho_{t}\right]|x'\rangle-\Gamma(x-x')\langle x|\hat\rho_{t}|x'\rangle,
\end{equation}
where $\hat \rho_t$ is the statistical operator of the system at time $t$, $\hat H$ is the system's Hamiltonian, and
\begin{figure}[b]
    \centering
  \includegraphics[width=0.8\linewidth]{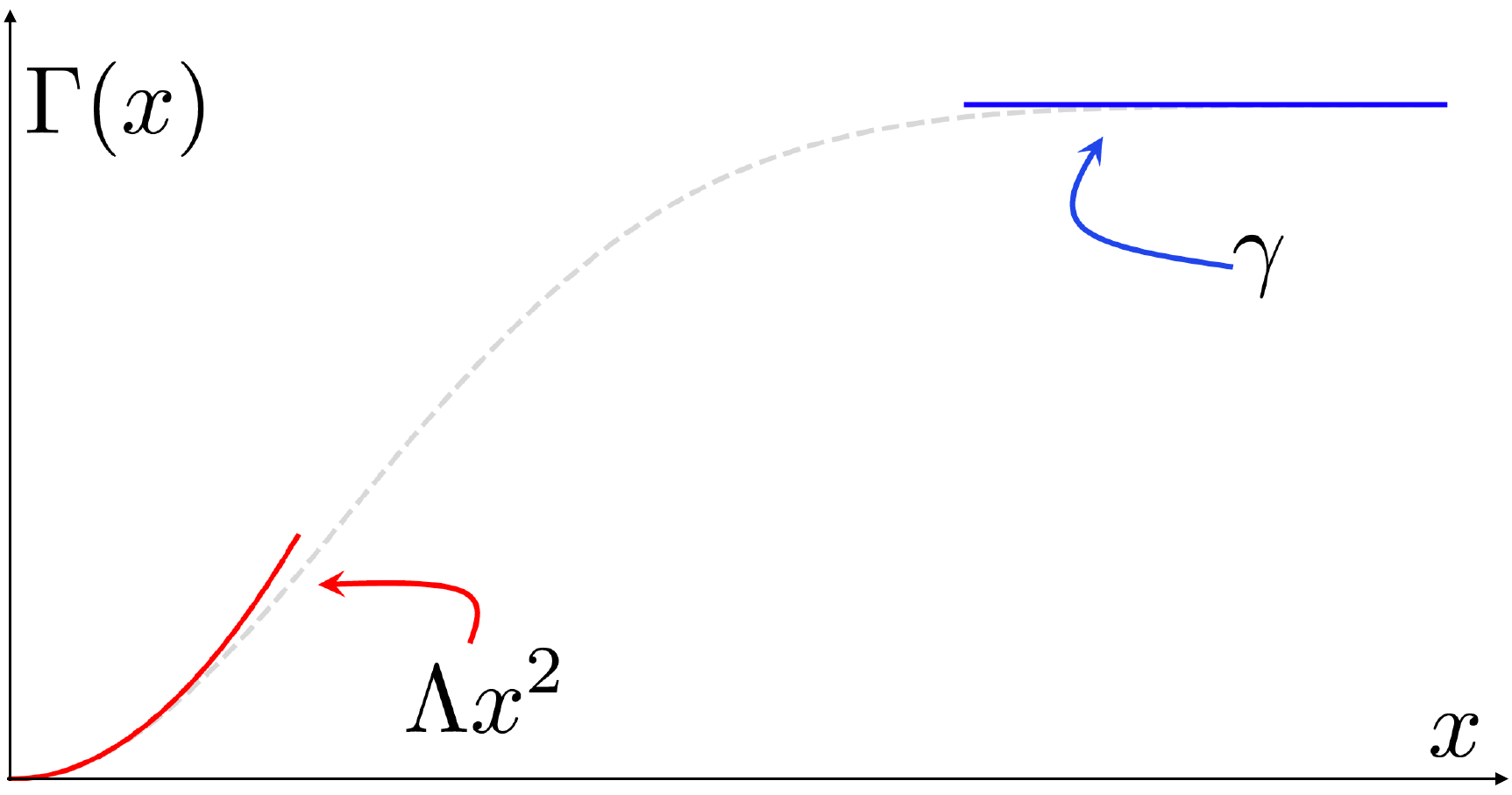}
    \caption{Typical dependence of the decoherence function $\Gamma(x)$ from the delocalization distance $x$. The relevant limits are $\Gamma(x)\sim\Lambda x^2$ for $x\ll a$ and $\Gamma(x)\sim\gamma$ for $x\gg a$, where $a$ is the characteristic length of the noise~\cite{PhysRevA.84.052121}.}
    \label{fig:decoherencefunction}
\end{figure}
the last term of Eq.~\eqref{master_gen} describes the deviations from  unitary dynamics occurring at a rate $\Gamma(x)$, which quantifies the decoherence effect. The typical behaviour of the latter, with a quadratic dependence for small spatial separations and saturating for large ones, is shown in Fig.~\ref{fig:decoherencefunction}. Such  deviations from unitarity can be due to environmental noises or non-standard modifications of quantum mechanics~\cite{RevModPhys.75.715,PhysRevA.84.052121,RevModPhys.85.471}. The environmental influence is always present, and it inevitably disturbs the experiment compromising the possibility to detect superposition states.  
The typical noise effects on the experimental setups addressed in this paper are due to collisions with residual gas particles, blackbody radiation, vibrations and in general any noise propagating through the experimental setup. 
{Quantitatively, for} a sphere made of fused silica with a radius of $60\,$nm and an internal temperature of $40\,$K, 
placed in vacuum in an environment at $20\,$K and a pressure of $10^{-11}\,$Pa { [cf.~Table~\ref{tab:parameterssmall}]}, {one has that for spatial superpositions larger than a nanometer but smaller than a millimeter, which is the range of interest for the interferometric test we will consider here, the }gas collisions give a constant contribution {$\Gamma(x-x')\sim1.1\,$s$^{-1}$} [cf. Ref. ~\citeonline{PhysRevA.84.052121}], while the contribution from blackbody radiation depends explicitly on the superposition size as $\Gamma(x-x')\sim 4.9\times 10^{12}\times |x-x'|^2$\,m$^{-2}$s$^{-1}$.

As mentioned, Eq.~\eqref{master_gen} is conducive of an investigation on potential deviations from standard quantum theory due, for instance, to collapse models. 
Due to its interesting phenomenology, theoretical interest and current strong experimental effort in testing it~\cite{RevModPhys.85.471,carlesso2019current,bassi2014collapse,vinante2019testing,carlesso2019collapse,vinante2020narrowing},  in the following we will focus on {the Continuous Spontaneous Localization (CSL) model}.
{The CSL model describes, through a stochastic and non-linear modification of the Schr\"{o}dinger equation, the collapse of the wave function as a spontaneous process whose strength increases with the mass of the system~\cite{Ghirardi1990a}. Its action is characterized by two phenomenological parameters: $\lambda_\text{CSL}$ and $r_c$. \MP{These are respectively the} collapse rate, \MP{which quantifies the strength of the collapse noise,} and the localization length of the model, \MP{setting the spatial resolution of the collapse}. Theoretical  considerations lead to different proposed values for such parameters: $\lambda_\text{CSL}=10^{-16}\,$s$^{-1}$ and $r_c=10^{-7}\,$m for Ghirardi, Rimini and Weber~\cite{ghirardi1986unified}; $\lambda_\text{CSL}=10^{-8\pm2}\,$s$^{-1}$ for $r_c=10^{-7}\,$m, and $\lambda_\text{CSL}=10^{-6\pm2}\,$s$^{-1}$ for $r_c=10^{-6}\,$m by Adler \cite{adler2007collapse}.}
{Consequently, one can describe the evolution of} the density matrix of a system 
{with} Eq.~\eqref{master_gen} where, in addition to the decoherence effects ascribed to the environment, a term accounting for spontaneous collapse appears. \MP{The form of such term and its effects are discussed in detail in Section \ref{Applications}.} This reveals the importance of a careful characterization of environmental sources of decoherence in view of probing new physics, which is the aim of the space experiments with large nanoparticles reviewed here. It should be noted indeed that, the experimental setups considered here are relevant also for testing other models  predicting non-standard decoherence mechanisms~\cite{bahrami2014schrodinger,pikovski2015universal,gasbarri2017gravity} or models like the Di\'osi-Penrose (DP) one~\cite{diosi1987universal,PhysRevA.40.1165,penrose1996gravity} in which the wave function collapse is related to gravity.

\subsection{Possible advantages of a space environment}\label{advantage}
\begin{figure}[t!]
\centering
\includegraphics[width=\columnwidth]{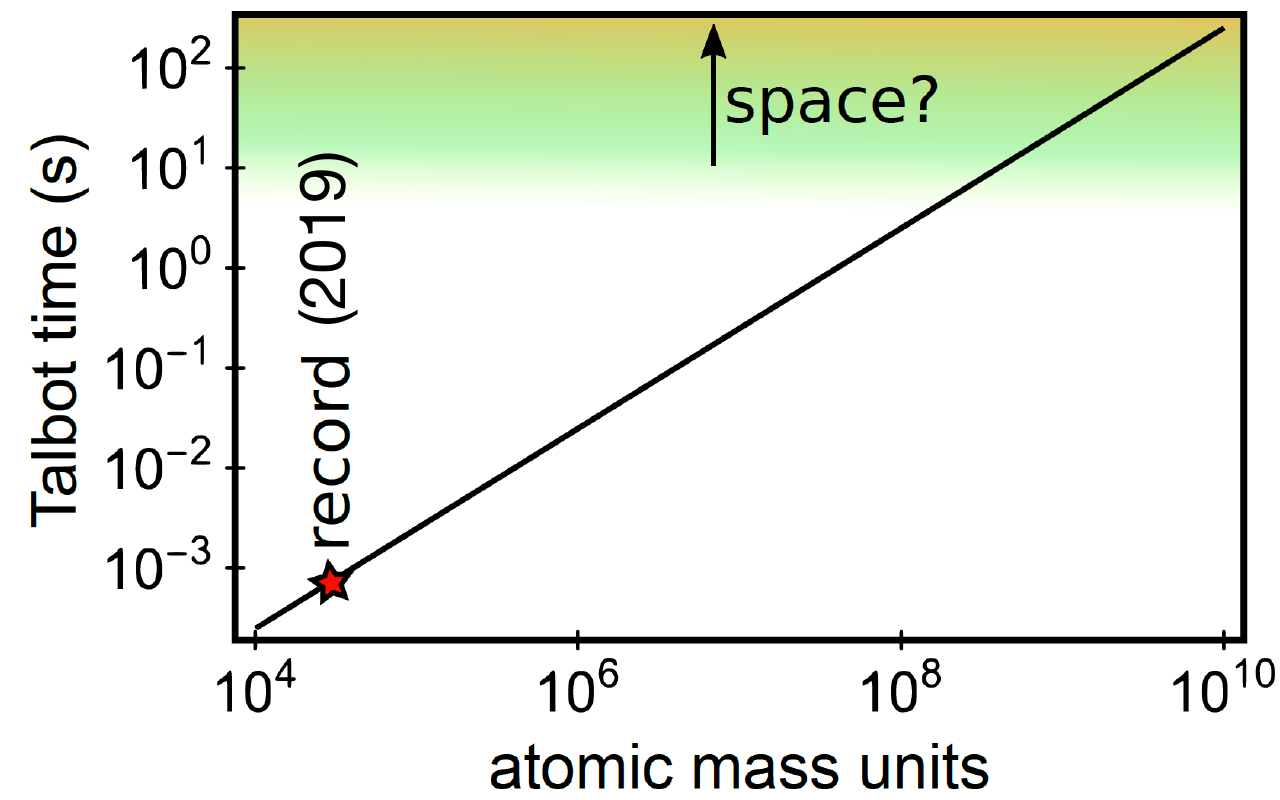}
\caption{The free-fall time for the test particles in a near-field interferometer is of the order of the Talbot time $t_T=m d^2/h$, where $m$ is the particle mass, $h$ is Planck's constant, and $d$ is the grating period. In experiments with significantly higher test masses than the current record~\cite{fein2019quantum}, the required free-fall time may eventually necessitates a space environment~\cite{kaltenbaek2016macroscopic}.}
\label{fig::talbotTime}
\end{figure}

The main advantage offered by space for quantum experiments with large particles is undoubtedly a long free-fall time. While freely-falling systems are not necessary for some non-interferometric experiments, they are the golden standard for the interferometric ones. For the latter, long free-fall times are of crucial importance to achieve better sensitivity and to increase the mass of the particles in quantum superposition \MP{as the rate of the wavefunction spreading is set by $1/m$.} In state-of-the-art interferometric experiments, and for masses of up to $10^6$\,amu, the necessary free-fall times are far below $1\,$s and can be readily achieved in laboratory experiments~\cite{bateman2014near,fein2019quantum,gasbarri2020prospects}. However, going to significantly higher test masses requires correspondingly longer free-fall times~\cite{kaltenbaek2012macroscopic,kaltenbaek2016macroscopic,gasbarri2020prospects} such as to eventually rendering it inevitable to perform such experiments in space (see Fig.~\ref{fig::talbotTime}).
Long free fall times help also in non-interferometric settings. The latter do not require the creation and verification of quantum superpositions but are based on the modified dynamics predicted by alternative models to quantum mechanics -- as for example the heating induced by the CSL noise on massive particles. In this context, letting the particle fall freely allows to reduce \MP{the effects of all the sources of noise that affect the centre of mass motion. Among them certainly is acceleration noise typically originating from mechanical vibrations. However, one should also include other forces acting on the particle's motion and which might be} present in the experiment thus maximizing the effects induced by modifications of quantum mechanics. \MP{In what follows, we provide a brief yet rigorous account of the most relevant of such forces.} 

An equally important challenge is the isolation from vibrations, 
which contribute to the overall decoherence mechanisms acting on the system. Especially in the low-frequency regime, space experiments can provide strong advantages compared to those performed on ground. {For example, ensuring that an interference pattern with a period of $d=1\,\mu$m formed during an evolution time of $T=100\,$s is not washed out requires a maximum acceleration noise of $S_{aa}^\text{max}\sim3d^2/8\pi T^3$ corresponding to $\sqrt{S_{aa}}\sim 3.5 \times 10^{-10}\,\mathrm{m\, s^{-2}/\sqrt{Hz}}$.}
Such low noise can be achieved in space. The most impressive achievement so far has been LISA Pathfinder with an acceleration noise as small as $\sqrt{S_{aa}}\sim 10^{-15}\,\mathrm{m\, s^{-2}/\sqrt{Hz}}$ in the mHz regime \cite{PhysRevLett.120.061101}. 
This value has to be compared to those of the state-of-art ground-based experiments. For example, the Bremen drop-tower allows for up to around 9\,s of free fall in an environment characterized by an acceleration noise of $\sim 10^{-5}\,\mathrm{m\, s^{-2}/\sqrt{Hz}}$ (cf. Ref.~\citeonline{Selig2010a}). 
\MP{We also mention that the exceptionally low level of noise achieved in LISA Pathfinder has already allowed this experiment to provide bounds on the CSL parameters \MP{that are} more than three orders of magnitude stronger than \MP{those} provided by the ground-based gravitational waves detector LIGO~\cite{carlesso2016experimental,helou2017lisa,carlesso2018non}, \MP{thus demonstrating the advantages -- in terms of isolation from vibration -- of space-based experiments}.}


A potential additional advantage of a dedicated space mission is data statistics. In state-of-the-art matter-wave experiments~\cite{fein2019quantum}, many test particles pass through the interferometer simultaneously. In proposals, considered in the QPPF, suggesting to prepare the initial state of the test particle by optomechanical means, the test particles pass through the interferometer individually~\cite{RomeroIsart2011b,kaltenbaek2012macroscopic,bateman2014near}. Thus, the time of the experiment is inevitably longer, and growing with the number of data points required. For example, in an experiment with $10^4$ data points, where a single shot takes $100\,$s ($10\,$s), the data collection would last more than $11.5$ days ($27$ hours). This compares very favourably to the typical number of two or three runs per day that can be performed in a micro-gravity environment on the ground as at the Bremen drop-tower, \MP{which is limited by the necessity of \MP{setting and resetting the pressure in} the entire tower between two consecutive drops} 
\footnote{\MP{It should be mentioned that this limitation is not present in the Hannover {\it Einstein Elevator} \MP{platform} where, for free-fall times of $\le 4\,$s, 300 runs per day are possible~\cite{lotz2020tests}.} }.

\subsection{State-of-the-art technological platforms}
The space environment promises, in principle, to provide a unique combination of low temperature, extremely high vacuum and very long free-fall times. In particular, the temperature in space is naturally limited by the temperature of the microwave background radiation of about $3\,$K while the vacuum is instead limited by the presence of cosmic and solar radiation~\cite{Biermann1957a}. However, in actual space-based experiments, additional shielding may be required. For example, if the spacecraft is in an orbit about one astronomical unit from the Sun, the payload will have to be shielded from direct solar radiation. {The spacecraft will require stabilization using microthrusters, which will introduce force noise, and there will need to be station-keeping maneuvers. These measures as well as the gravitational field of the spacecraft will reduce the achievable free-fall time.} In addition, the equipment necessary to operate the payload and the spacecraft typically will be in an enclosure kept at a stable temperature of about $300\,$K. Inside the spacecraft, it is therefore not trivial to achieve cryogenic temperatures and extremely high vacuum levels. Achieving the vacuum levels and temperatures necessary for macroscopic tests of quantum physics therefore requires careful considerations. In the context of the MAQRO mission concept, it has been suggested to use purely passive radiative cooling and direct outgassing to space in order to achieve these requirements~\cite{kaltenbaek2012macroscopic,kaltenbaek2016macroscopic,hechenblaikner2014cold,zanoni2016thermal}. During the QPPF study, this concept was adapted protect the scientific instrument with a cover and to enhance the cooling performance ny additional active cooling using a hydrogen sorption cooler~\cite{QPPF}. 

The protective cover limits outgassing to space, and the QPPF study concluded that the achievable pressure would at best be
$10^{-11}$\,Pa instead of the aimed pressure of $10^{-13}$\,Pa. As a result, 
the experiments were constrained to a test particle mass up to $2\times10^9$\,amu and free-fall time up to $40\,$s. Because this has a significant impact on the science objectives, improving the achievable vacuum in space-based experiments will be a critical issue to be solved before a space mission of this type can be launched. Two other critical issues were identified in the QPPF study~\cite{QPPF}. 
{Firstly, the mechanism needed for loading the test particles into an optical trap in extremely high vacuum conditions needs further scrutiny and several viable alternatives are under consideration. The QPPF study suggested to desorb particles from microelectromechanical systems (MEMS) and guide them to the experiment using linear Paul traps~\cite{QPPF}, a mechanism that is currently under further investigation by ESA. An alternative suggestion makes use of a combination of linear Paul traps and hollow-core photonic-crystal fibers~\cite{kaltenbaek2016macroscopic} or the desorption of particles from a piezoelectric substrate using surface acoustic waves.  
The challenge of such desorption-based approaches is to make sure that the desorbed sub-micron particles do not carry a net charge~\cite{PhysRevA.95.061801}, and that their center-of-mass motion is sufficiently cold to allow for optical trapping. At the same time, a sufficiently low internal temperature of the particles is required to avoid decoherence due to the emission of blackbody radiation~\cite{RomeroIsart2011b,kaltenbaek2012macroscopic}. Secondly, the optical gratings used for preparing non-classical states has grating apertures comparable to the size of the nanoparticles to be employed. This can decohere the quantum states via photon scattering. A recent study~\cite{PhysRevA.100.033813} investigated the latter issue extending the formalism of near-field interferometry beyond the point-particle approximation and offering the basis for the analysis reported in Sec.~\ref{ProofOfPrincipleImlementation}.}


\section{Non-interferometric tests}\label{Applications}

In this section, we focus on non-interferometric tests of quantum mechanics. Differently from the interferometric ones, this class of tests does not rely on the availability of quantum superpositions but are based on side-effects of modifications of quantum mechanics. {Consequently, they can be performed also in presence of strong decoherence, although the latter will influence the effectiveness of the test. For this reason, they currently provide the most stringent tests of collapse models on ground. 
}

A plethora of different experiments belongs to this class and exploit different physical systems. {Among them, precision measurements of the internal energy of a solid, expected to vary due to the collapse noise, have been exploited in Refs.~[\citeonline{adler2018bulk,bahrami2018testing,adler2019testing}]. The modifications to the free evolution dynamics of Bose-Einstein condensate due to the presence of the collapse mechanism has been investigated in Refs.~[\citeonline{laloe2014heating,bilardello2016bounds}]. And X-ray measurements -- which exploit the fact that the collapse mechanism makes charged particles to emit radiation~\cite{fu1997spontaneous,curceanu2016spontaneously,piscicchia2017csl} -- have already provided strong limits on the Di\'osi-Penrose model~\cite{donadi2020underground}. In this context, also optomechanical experiments are of particular relevance~\cite{vinante2016upper,carlesso2016experimental,PhysRevA.98.022122,vinante2017improved,helou2017lisa,zheng2020room,vinante2020narrowing,pontin2020ultranarrow}. They are typically used to characterize noise~\cite{aspelmeyer2014cavity,millen2020optomechanics,goldwater2019quantum}, and thus possibly discriminate between standard and non-standard noise sources~\cite{vinante2020narrowing}.}

{One of the most promising non-interferometric test in space is based on monitoring the expansion of the centre-of-mass position spread of a freely-falling nanoparticle~\cite{PhysRevA.93.062102}. The main reason, as it is shown in Eq. (\ref{variance}) below, is that the position variance grows as the cube of time, making evident the advantage of the long free-fall time that can be achieved in space.
It could be argued that long times can also be achieved in ground experiments by suspending the particles using an harmonic trap. However, the use of such a trap would certainly introduce additional noises and, more importantly, it would imply a position variance growth that scales only linearly with time~\cite{bilardello2016bounds,PhysRevA.94.010104}. 

Given the evolution in Eq.~(\ref{master_gen}), it is easy to show that its non-unitary part does not affect the average position $\langle x_{t}\rangle$ of the particle, but changes its variance $\sigma^{2}= \braket{x_{t}^{2}}-\braket{x_{t}}^{2} $ by a factor $\braket{\Delta \sigma^{2}}$ that, for a free system and in the $x\ll{a}$ regime [cf. Fig.~\ref{fig:decoherencefunction}], reads
\begin{equation}\label{variance}
\braket{\Delta \sigma^{2}}=\frac{2\Lambda\hslash^{2}t^{3}}{3m^{2}}.
\end{equation}
The diffusion rate $\Lambda$ is the sum of different contributions stemming from residual gas collisions,  blackbody radiation and non-standard sources, such as  the CSL or the Di\'osi-Penrose model.
For the CSL model and a homogeneous sphere of radius $R$ and mass $M$, one has~\cite{zheng2020room}
\begin{equation}
\Lambda_{\text{CSL}}=\frac{6 \lambda_\text{CSL}  M^2 }{m_0^2 R^2\eta^4_{CSL}}\left[\left(1+\frac{\eta^2_{CSL}}{2}\right) e^{-\eta^2_{CSL}}+\frac{\eta^2_{CSL}}{2}-1\right],
\end{equation}
while for the DP model one obtains
\begin{equation}
\begin{aligned}
\Lambda_\text{DP}&=\frac{M^{2}G}{2\hslash \sqrt{\pi}R^{3}}\left[\sqrt{\pi}\operatorname{erf}\left(\eta_{DP}\right)-\frac{3}{\eta_{DP}}+\frac{2}{\eta^{3}_{DP}}\right.\\
&\left.+\frac{e^{-\eta^2_{DP}}}{\eta_{DP}}\left(1-\frac{2}{\eta^{2}_{DP}}\right)\right].
\end{aligned}
\end{equation}
We have used the dimensionless parameters $\eta_{CSL}=R/r_c$ and $\eta_{DP}=R/R_0$ with  $R_0$ a free parameter that is characteristic of the DP model~\cite{bahrami2014role}. These expressions can be then used to set bounds on, respectively, CSL and DP parameters with space-based experiments, as we discuss next.
}

\subsection{Long free-fall times: opportunities and challenges for space-based experiments}
{A possible space-based experiment, as envisioned in the MAQRO proposal and QPPF, is as follows. A nanosphere is initially trapped by an harmonic optical potential and its center-of-mass motion is optically cooled. The trapping is then removed and the nanosphere remains in free-fall for a time $t$ after which its position is measured. Achieving a high position resolution is possible by, for example, combining a coarse-grained standard optical detection on a CMOS chip with a high-resolution backscattering detection scheme~\cite{Tebbenjohanns2019a}, which could eventually provide a position accuracy on the order of $\varepsilon=10^{-12}\,$m \MP{at a typical bandwidth of 100 kHz, by controlling the measurement back-action.~\cite{vovrosh2017parametric}}
By repeating such a procedure $N$ times, one can reconstruct the position spread $\sigma^{2}$ and thus quantify the effects of the non-unitary dynamics through Eq.~(\ref{variance}). To detect effects as those predicted by the CSL or the DP model, one needs to minimize the competing standard decoherence effects (from collisions and blackbody radiation), which contribute to the total $\Lambda$ in Eq.~(\ref{variance}). 
\begin{figure}[t]
    \centering
    \includegraphics[width=\linewidth]{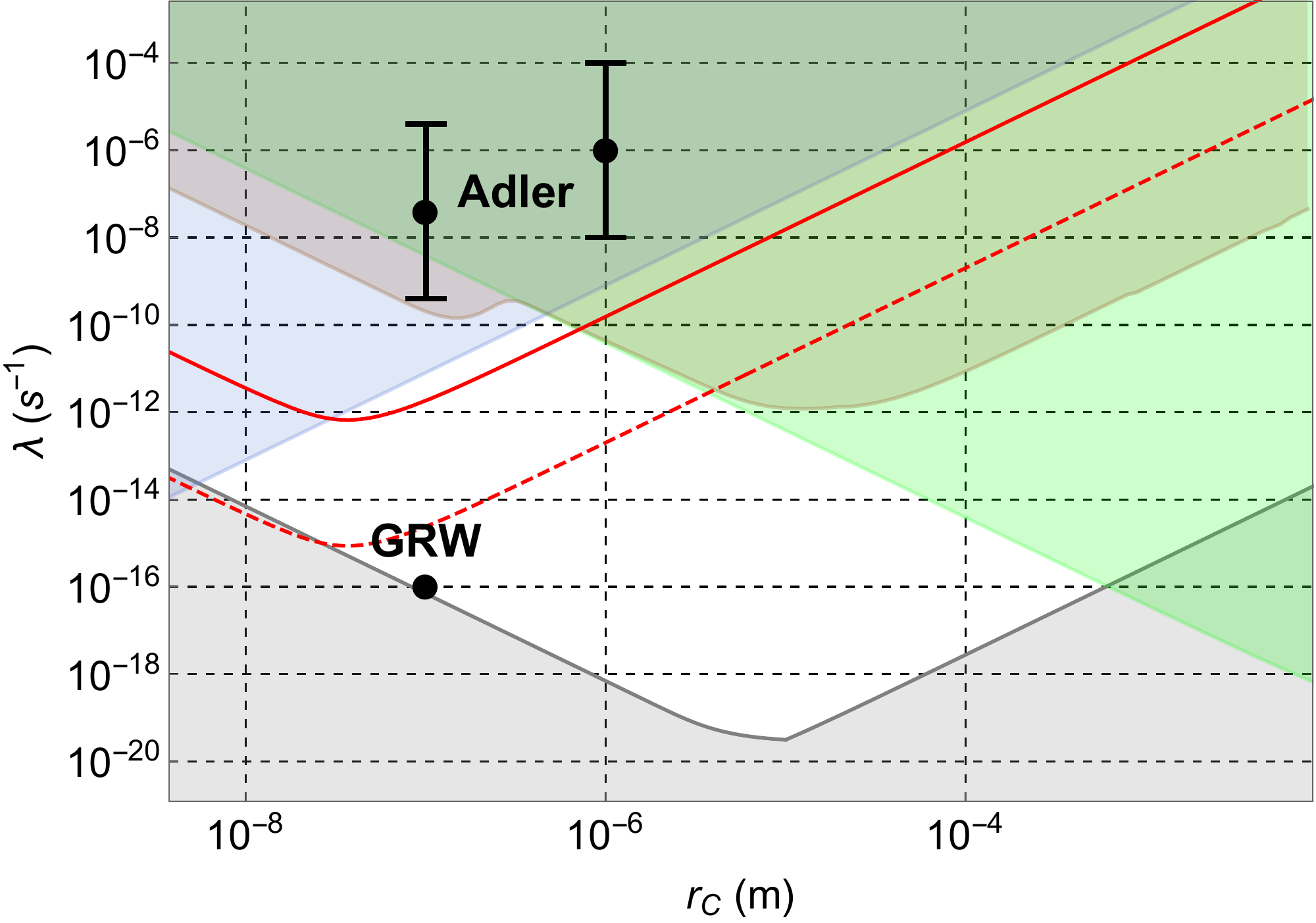}
    \caption{Exclusion plots for the CSL parameters $\{r_c,\lambda\}$ from non-interferometric experiments. \MP{The solid, red line represents the bound on the CSL parameters that can be \MP{potentially} achieved through non-interferometric experiments in space with the parameters in Table~\ref{tab:parameterssmall}. Here, the main limitation is due to the environmental conditions of pressure and temperature. The dashed red line indicates the upper bound that could be obtained by decreasing the pressure to $P=3 \times 10^{-14}$~Pa, so that the main limitation would be represented by the statistical error. These bounds are} compared to the strongest bounds and corresponding excluded parameter regions present in literature: X-rays emission (blue region)~\cite{piscicchia2017csl}, LISA Pathfinder (green region) \cite{carlesso2016experimental,carlesso2018non}, multilayer cantilever (brown region) \cite{vinante2020narrowing}. The gray region is the theoretical lower bound, \MP{which is estimated by requiring the collapse to become effective at the mesoscopic scale where the quantum-to-classical transition is expected}~\cite{torovs2017colored}.   }
    \label{fig:noninterf}
\end{figure}
{We are now in the position to estimate the bounds on the CSL parameters.} To do this,
we employ the values in Table~\ref{tab:parameterssmall}. We consider silica nanospheres with a 120\,nm diameter as test particles and an
internal temperature fixed at 40\,K. 
Moreover, we also assume levels of vibrational noise similar to those obtained in LISA Pathfinder~\cite{PhysRevLett.120.061101}. With these assumptions, the strongest competing effect to the CSL noise is the collisional decoherence, which limits the bounds on the CSL parameters.  Such a bound is indeed obtained by setting $\Lambda_\text{CSL}$ equal to the collisional contribution to the diffusive constant $\Lambda$, whose form is reported in Ref.~[\citeonline{PhysRevA.84.052121}].} \MP{We show  the corresponding bound as the solid red line in Fig.~\ref{fig:noninterf}, where such bound is compared to ground-based ones achieved by  state-of-the-art experiments on the CSL model.}

For what concerns the DP model, the state-of-the-art experimental bounds indicate that the free parameter $R_0$ is limited to~\cite{donadi2020underground} $R_0\geq R_0^*= 0.5 \times 10^{-10}$\,m. Because the DP-induced collapse becomes stronger for smaller $R_0$, the maximum effect is obtained for~{$R_0^*$.} Such a value of $R_0$ leads to a position spread in the aforementioned set-up of $\sqrt{\braket{\Delta\sigma^2}}\sim3\times 10^{-26}\,$m {for $t=100\,$s}, well beyond the state-of-art position measurement sensitivity $\varepsilon$.

An important aspect to consider for experiments performed in space is their limited lifetime. Especially when one considers long free-fall times, this will have an impact on the statistical accuracy with which one can determine the variance of the measured data points~\cite{kaltenbaek2016macroscopic,QPPF,Kaltenbaek2021a}. The long free-fall time $t$ required to see potential deviations from the quantum predictions has to compete with a finite time $T$ available to take the complete data set. At best, the number of data points can be $N=T/t$.
{This limit on the number of data points implies a statistical uncertainty in determining the position spread. To quantify it, we assume that the initial quantum state of the test particle is the ground state of an harmonic oscillator with a mechanical frequency $\omega$. Consequently, the measured position will be normally distributed and the corresponding fractional uncertainty of the variance of the measured position will be~\cite{Taylor1997a} $\sqrt{2/(N-1)}\approx \sqrt{2 t/T}$.}
{Assuming} the deviations from the quantum predictions to be small, the statistical uncertainty 
of the variance is $\Delta x^2_f\approx \sqrt{2 t/T} x^2_s$, where $x^2_s\approx  t^2\omega\hslash/2m$ is the variance of the wavepacket predicted by quantum physics for times much longer than $1/\omega$. 
{By taking} $\omega=10^{5}\,$Hz for the trap frequency, a free evolution time $t=100\,$s and a total time $T$ of 30 days, we have that $\Delta x_f\sim 3\times 10^{-5}\,$m which has to be compared to the sensitivity $\varepsilon\sim10^{-12}\,$m. For these parameters, the statistical uncertainty will dominate over the position sensitivity $\varepsilon$ already after about~\cite{Kaltenbaek2021a} $0.1\,$ms. 
{Such a statistical uncertainty becomes a fundamental limitation for the experiment. The corresponding upper bound on the CSL parameters is represented by the dashed, red line in Fig.~\ref{fig:noninterf}. To reach such a limit, the pressure would need to be reduced by more than two orders of magnitude, down to $P=3 \times 10^{-14}$\,Pa,  with respect to the conditions setting the continuous red bound. }

\MP{Fig.~\ref{fig:noninterf} and the analysis above \MP{suggest} that non-interferometric experiments performed with the parameters in Tab.~\ref{tab:parameterssmall} can enhance only partially the exploration of the CSL parameter space. A more substantial improvement would require to solve technical challenges, \MP{such} as a significant pressure reduction. Alternatively, one can pursue the path of interferometric experiments, which is discussed in the next section.}


\section{Interferometric tests}\label{ProofOfPrincipleImlementation}

Here, we will provide an overview of the current state-of-the-art for proposals of interferometric experiments testing the superposition principle of quantum mechanics for higher masses than the current experimental record on ground by using a space environment. We will discuss the challenges faced by such experiments, and we will provide novel simulations results estimating the interference visibility expected in space-based experimental tests of the superposition principle of quantum mechanics. 

\subsection{Near-field interferometry}

After Clauser envisioning its use for \textit{small rocks and live viruses} experiments~\cite{clauser1997broglie} and its initial demonstration for C$^{70}$ molecules interferometry~\cite{PhysRevLett.88.100404}, for almost two decades the most successful technique harnessed for interferometric tests of quantum physics has been near-field interferometry~\cite{RevModPhys.84.157,arndt2014testing}.
With this technique, and employing three optical gratings~\cite{gerlich2007kapitza}, in 2019 the Arndt's group in Vienna was able to successfully build and demonstrate spatial the quantum superposition of big molecules with masses beyond~\cite{fein2019quantum} $10^4$\,amu.

Recently, the possibility to consistently describe the effects of an optical grating on large dielectric particles with radii comparable to the optical wavelengths~\cite{Nimmrichter2014a,PhysRevA.100.033813} has opened the possibility to use optical grating to study quantum interference on even larger particles. At the same time,
concrete proposals to go beyond the current mass record, employing individually addressed dielectric particles and single optical grating~\cite{bateman2014near,kaltenbaek2016macroscopic,QPPF} have shown the experimental viability of near-field interferometry to actually perform such larger mass superposition experiments.

We thus focus specifically on these implementations to give a overview of how a near-field interferometric scheme works.
\begin{figure}[t!]
\centering
\includegraphics[width=\columnwidth]{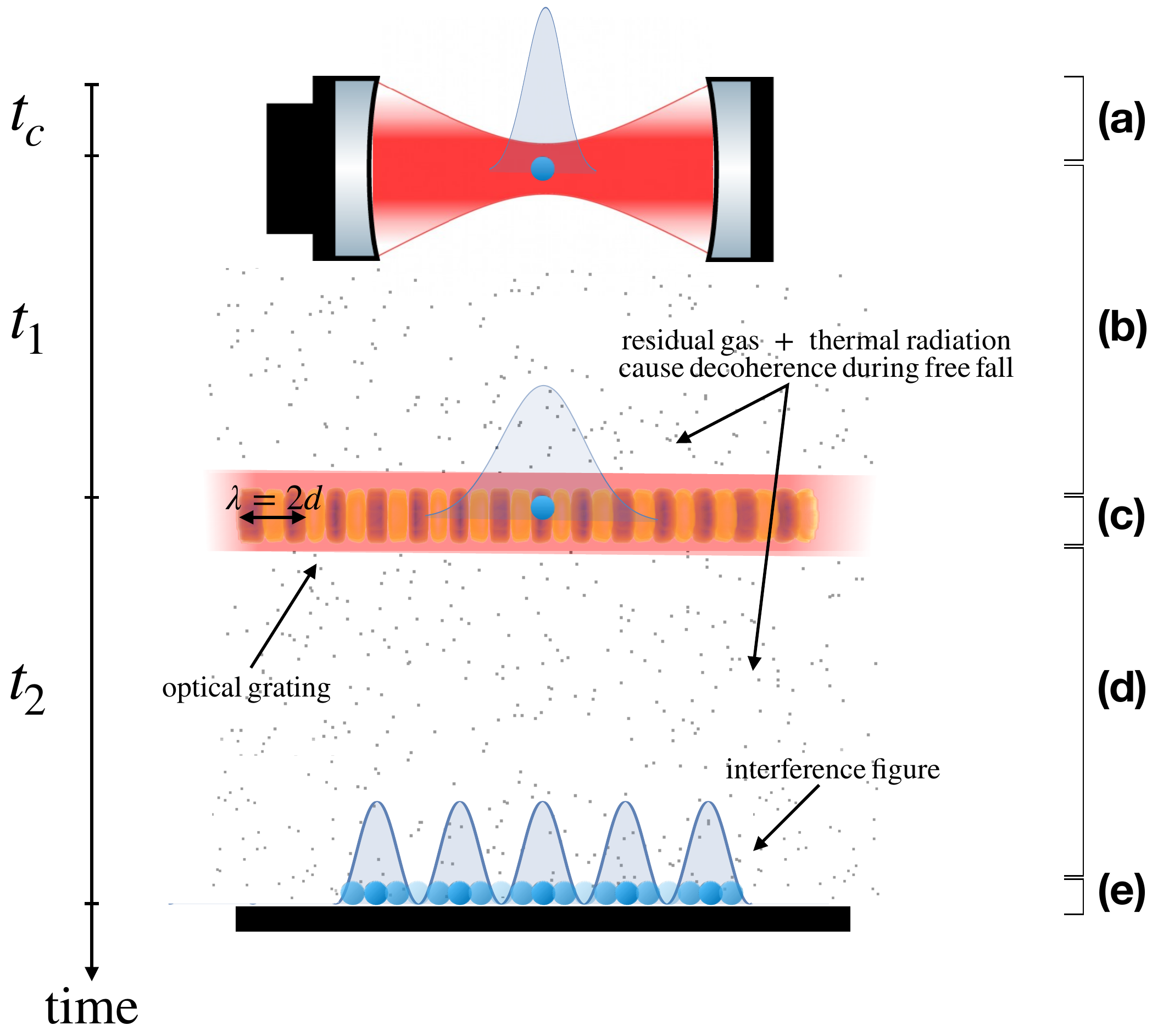}
\caption{{
Schematic representation of a space-based near-field interferometry experiments referred to in the text. While for ground-based experiments the time axis corresponds also with the vertical position of the particle, in space-based experiments the particle remains in the same position (relative to the experimental apparatus) and the trapping laser, the grating laser and a final measurement laser must be activated in turns at different times.}}
\label{figInonI}
\end{figure}
We refer to Fig.~\ref{figInonI} for a schematic representation of a single-grating near-field set-up. Contrary to the case of lighter systems, where molecular beams are engineered, each nanoparticle in the experiment is individually addressed. We thus have, at each run of the experiment, four main stages: 

\begin{itemize}
    \item[\textbf{(a)}] The nanoparticle is trapped and cooled down in an optical cavity for a time $t_c$ after which the center-of-mass degree of freedom is in a very low-temperature thermal state characterised by the momentum and position variances $\sigma_p,\,\sigma_z$. No cooling down to the ground state is required. 
    \item[\textbf{(b)}] The particle is released and freely falls for a time $t_1$. 
    During this time, residual gas collisions and thermal radiation are the main sources of decoherence. The free evolution of the post-cooling state needs to guarantee that the coherence-length is sufficient to cover at least two adjacent ``slits'' of the optical grating.
    \item[\textbf{(c)}] A retro-reflected pulsed laser provides a pure-phase grating~\cite{Nimmrichter2014a} for the dielectric nanoparticle. Scattering and absorption of grating photons constitute the main decoherence channels in the short interaction time with the grating.
    \item[\textbf{(d)}] Second period of free evolution for a time $t_2$ during which the same sources of decoherence as in point (b) act. This stage has to last enough time for the interference pattern to form.
    \item[\textbf{(e)}] The position of the particle is measured via optical detection~\cite{kaltenbaek2016macroscopic,QPPF}. 
\end{itemize}
By repeating this protocol \textbf{(a-e)} many times, an interference pattern can form in the measured position distribution. This pattern can be mathematically described by a probability distribution function $P(z)$ which can be analytically derived from a phase-space treatment of the interferometric experiment~\cite{bateman2014near,Nimmrichter2014a}:
\begin{equation}
\label{pattern}
\frac{P\left(z\right)}{\delta}= 
1+2\sum_{n=0}^{\infty} R_{n} \,B_{n} \left[\frac{n dt_{2}}{t_{T}D}\right] \cos\left(\frac{2 \pi n z}{D}\right)e^{-2\left(\frac{n\pi\sigma_{z}t_{2}}{Dt_1}\right)^2},
 \end{equation}
where $\delta=m/\left(\sqrt{2 \pi} \sigma_{p}(t_{1}+t_{2})\right)$, $t_T= md^2/h$ is the Talbot time and $D=d(t_1+t_2)/t_1$ is a geometric magnification factor. In this last expression, the $B_{n}$'s are known as the generalized Talbot coefficients~\cite{Nimmrichter2014a,PhysRevA.70.053608} and account for the coherent and incoherent effects of the optical grating, while the kernels $R_{n}$ account for environmental decoherence, due to absorption, emission, and scattering of thermal radiation and collisions with residual gas, during the free-falling times $t_1, t_2$. 
This expression remains formally unchanged when classical particles following ballistic trajectories are considered, but the explicit expressions for the decoherence kernels and Talbot coefficients will change. 
Expression \eqref{pattern}, with the proper coefficients, can thus be used to describe the classical shadow pattern arising  from a completely classical description of the system (see Fig.~\ref{fig:carpet}). 
Finally, non-linear modifications of quantum mechanics -- but  also other sources of positional decoherence like e.g. a stochastic gravitational waves background~\cite{bassi2017gravitational,asprea2019gravitational} -- can easily be included in Eq.~\eqref{pattern} by introducing their respective noise kernels $R_n$. We refer the interested reader to Refs.~[\citeonline{PhysRevA.100.033813,gasbarri2020prospects}] and the Supplemental Material (SM)~\cite{SI} for a detailed derivation and explicit expressions of the functions entering Eq.~\eqref{pattern}.

In order to go beyond the current near-field interferometry mass record, large particles need to be used. Here, ``large'' refers to spherical particles with a radius $R$ comparable to or greater than the grating period $d$ such that $kR\gtrsim 1$, where $k=2\pi/\lambda$ is the wave-vector of the optical grating. In the following, we will use the formalism developed in Ref.~[\citeonline{PhysRevA.100.033813}], and based on Mie scattering theory, to account for a large particle traversing an optical grating. For what concerns the pure-phase character of the grating -- i.e., its coherent effect on the particle's state -- it can be shown that the unitary evolution of the particle's state $\hat{\rho}$ (reduced along the longitudinal direction $z$) when traversing the grating assumes, in the eikonal approximation, the form  $\bra{z}\rho\ket{z'}\rightarrow \exp\left[{-i\phi_0 (\cos^2{kz}-\cos^2{kz'})}\right]\bra{z}\rho\ket{z'},$
where $\phi_0$ is the eikonal phase factor characterising the coherent evolution. This is the same as in the case of a point-like particle and the only difference introduced by the use of Mie scattering theory~\cite{mie1908beitrage,bohren2008absorption} is found in the structure of the eikonal phase $\phi_0$ which can be expressed as 
\begin{equation}\label{phi0mie} 
    \phi_0=\frac{8F_0 E_L}{\hslash c \epsilon_0 a_L k|E_0|^2},
\end{equation}
in terms of the laser and particle's parameters. Here, $c\epsilon_0|E_0|^2/2$ is the intensity parameter of the incident light, $E_L$ and $a_L$ are the grating laser energy and spot area respectively, and $F_0$ is obtained from Mie theory upon the evaluation at $z=-\lambda/8$ of the longitudinal conservative force acting on the particle (see Refs.~[\citeonline{Nimmrichter2014a,PhysRevA.100.033813}] and references therein). Eq.~\eqref{phi0mie} reduces to the well-known result $\phi_0=2 \mathcal{R}(\chi)E_L/(\hslash c\epsilon_0 a_L)$ with $\chi$ the polarizability for a point-like particle. For what concerns the incoherent effects of the grating, the finite-size of the particles leads to modify the Talbot coefficients with respect to the point-like case. We refer the reader to Ref.~[\citeonline{PhysRevA.100.033813}] and the SM~\cite{SI} for further details.

\begin{figure}
 \centering
\includegraphics[width=1.\columnwidth]{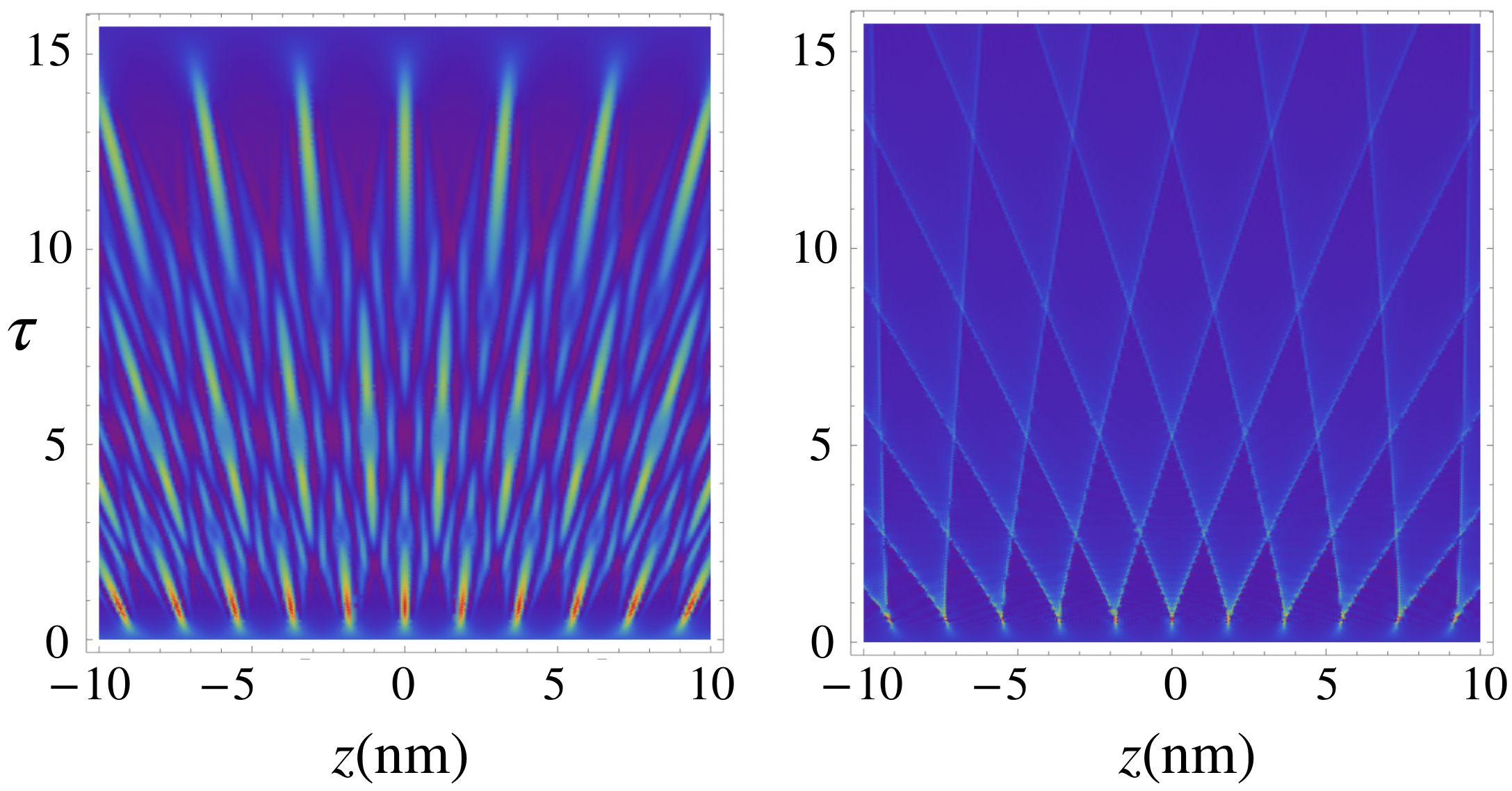}
\caption{Talbot carpet arising from pure phase-grating without any source of decoherence for a point-like particle. This picture shows the comparison between the interference pattern predicted by quantum mechanics, $P(z)$ in Eq.~\eqref{pattern} (on the left), and the shadow pattern that is formed by classical particles following ballistic trajectories (on the right) for different values of $\tau=t_2/t_T$. It is then apparent that, in order to can claim the observation of a quantum superposition, we need to be able to distinguish between these two patterns.}\label{fig:carpet}
\end{figure}

Finally, as discussed in Ref.~[\citeonline{gasbarri2020prospects}], large particles near-field interferometric experiments presents several technical challenges. Common to both ground and space-based experiments is the challenge of diminishing as far as possible any environmental noise which would suppress the interference pattern. This can be achieved by a combination of ultra-high vacuum and cryogenic conditions. Moreover, for experiments aiming at using single particles in several ($\sim 10^4$) runs, fast reloading/recycling technique must be developed~\cite{grass2016optical,mestres2015cooling,PhysRevLett.121.063602,QPPF}.
On top of these challenges, the key limitation for ground-based experiments is the short free-fall time. This is due to the Earth gravitational field and limits such experiments to a few seconds of free evolution. While this challenge can be overcome in principle, it will require a substantial modification of the scheme to go beyond masses of the order of $10^7$~amu~\cite{pino2018chip,gasbarri2020prospects}. This is not the case for space-based experiments, where current estimates show the promise to reach masses of the order of $10^9-10^{11}$~amu and free-falling times of the order of hundreds of seconds~\cite{kaltenbaek2016macroscopic,QPPF}. In the following section, we substantiate these claims by presenting an optimized analysis of space-based near-field interferometry showing the actual possibilities offered by a space environment.  

\begin{figure*}[t!]
\centering
\includegraphics[width=1.\textwidth]{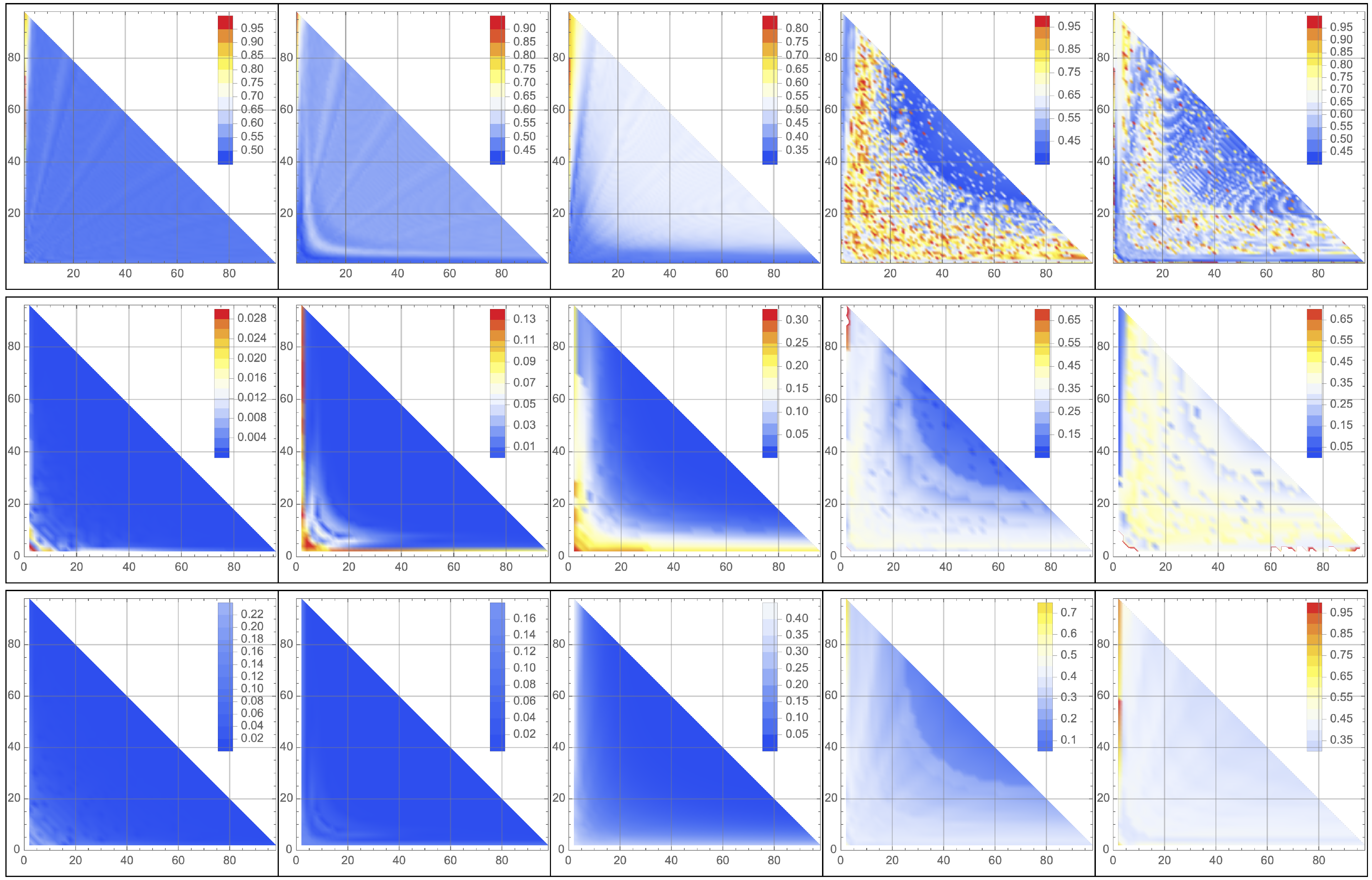}
\caption{Values of the cost functions $\aleph_{QC}$ and $\aleph_{QCSL}$ as a function of $t_1, t_2$. The triangular area of the density plot is determined by the constraint $t_1+t_2\leq 100$~s. The different columns correspond to five different values of the mass of the nanoparticles considered, respectively $10^{7},10^{8},10^{9},10^{10},10^{11}$~amu. The first row shows $\aleph_{QC}$, i.e., how distinguishable the quantum and classical interference figures are. These figures are obtained by choosing as $E_L/a_L$ the value that maximizes $\aleph_{QC}$ in the physically feasible range between $10^{-6}$ and $5$~J/m$^{2}$. For the numerical values of $E_L/a_L$ employed we refer to Fig.~2 in the SM~\cite{SI}. The second row shows the values of $\aleph_{QCSL}$, i.e.,  how distinguishable the quantum interference figure is from the one accounting for the CSL. These figures are obtained by assuming the same values of $E_L/a_L$ used in the first row and for a value of the CSL parameters proposed by Adler and given by $\lambda_\text{CSL} =10^{-8}$~s$^{-1}$ and $r_c=10^{-7}$~m. Finally, the third row shows the values of $\aleph_{QCSL}$ like in the second row where, however, the values of $E_L/a_L$ used are the ones that maximize $\aleph_{QCSL}$ independently from the results in the first row of figures. For the numerical values of $E_L/a_L$ employed we refer the interested reader to Fig.~1 in the SM~\cite{SI}. }\label{fig:QMCSLYet}
\end{figure*}
\subsection{Optimization for large particles: the current frontiers}
We present in this section the results of a numerical investigation of the possibilities offered by space-based experiments in conjunction with near-field interferomerty as discussed in the previous section. We employ the formalism developed in Ref.~[\citeonline{gasbarri2020prospects}] to account for the finite size of the particles with respect to the grating period, and we use the experimental parameters, as summarized in the SM~\cite{SI}, which have been extracted from the QPPF study about the MAQRO mission~\cite{QPPF}. We are able to include in our analysis all the major known sources of environmental decoherence which can affect the interference pattern. In particular, we account for scattering and absorption of grating photons at stage \textbf{(c)} of the protocol, residual gas collisions and black-body thermal radiation decoherence during the free evolution stage \textbf{(b)-(d)}. We then include the effect of modifications of quantum mechanics in the from of the Continuous Spontaneous Localization (CSL) model with white-noise~\cite{RevModPhys.85.471}. What we present here is the first fully consistent analysis of such a set-up and its potential for fundamental physics studies, which does not rely on the Rayleigh approximation, which cannot be consistently used unless for order of magnitude estimates.  

One complication of near-field interferometry -- in contrast with the textbook case of far-field interferometry -- that needs to be taken into account when performing an experiment is that also perfectly classical particles following ballistic trajectories through the optical grating would form a typical interference-like figure known as Moir\'{e} shadow pattern~\cite{RevModPhys.84.157} (see Fig.~\ref{fig:carpet}). It is thus of crucial importance to discriminate between this pattern and a quantum mechanical one~\cite{PhysRevLett.88.100404}. This is a prerequisite for both claiming to be able to test the superposition principle and for any analysis of modifications of standard quantum mechanics. Thus, we introduce a first figure of merit ($\aleph_{QC}$) to estimate the ``distance'' between the quantum interference patter's probability distribution (pdf) and the pdf of the shadow pattern which would result from classical mechanics
\begin{align}\label{alQC}
\aleph_{QC}= \frac{1}{L}\int_{-L/2}^{L/2}  \frac{|P_{\rm Q}(z)-P_{\rm Clas}(z)|}{|P_{\rm Q}(z)+P_{\rm Clas}(z)|} dz
\end{align}
where $L = 10^{-7}$~m is the spatial window in which the position measurement is performed and $P_{\rm Clas}$ ($P_{\rm Q}$) is the pdf predicted by classical (quantum) mechanics. A similar quantity can then be obtained to discriminate between a quantum interference pattern and the pattern deriving from  modifications of quantum mechanics. We will focus here on the CSL model with white noise. Thus the second figure of merit that  we will employ is $\aleph_{QCSL}$, which is given by Eq.~\eqref{alQC} with $P_{{\rm Clas}}\rightarrow P_{{\rm CSL}}$. 

In the following we assume to be able to discriminate values of $\aleph\ge 0.05$ (i.e., difference bigger than $5\%$) which appears to be an experimentally justifiable choice~\cite{juffmann2013experimental}. Moreover, we optimize over the parameters $t_1,t_2, E_L/a_L$ of the set-up, which can be easily controlled, to maximize the figure of merits. As we will see, the optimization leads to values of the figure of merits well above the $5\%$ threshold. Before illustrating the results of the analysis, let us comment on the choice of parameters. On the one hand, the free-falling times $t_1,t_2$ are extremely important in the formation of the interference figure, whether it is $t_1$ which guarantees a sufficient spreading of the initial state to a coherence length covering more than two ``slits'' or $t_2$ which allows the intereference to happen. These two times can also be easily adjusted in a space-based experiment by simply changing the activation times of the grating and measurement lasers. On the other hand, the parameter $E_L/a_L$ enters directly in Eq.~\eqref{phi0mie} and thus determines the pure-phase coherent effect of the grating. This parameter can also be easily tuned, being a property of the way the grating laser is operated. We keep instead fix all the other parameters entering our analysis (see the table reported in the SM~\cite{SI}). These are: the wavelength of the grating laser, which is dictated by current technological possibilities; the material(s) parameters of the nanosphere, we considered silica (SiO2) particles which are widely employed in optomechanical experiments for their optical properties; environmental parameters, which have been extracted from the QPPF study~\cite{QPPF} and represent the current state-of-the-art for space-based set-ups. Furthermore, always referring to the QPPF study on the stability of a possible mission's spacecraft, we constrain the total free fall time to $t_1+t_2\leq 100$~s. Note that, the QPPF study concludes that, due to vacuum restriction, the interference pattern for the proposed MAQRO mission would be visible for free-falling times of up to 40~s. However, the 100~s benchmark is among the scientific objectives of the community, as reported in the QPPF. We thus chose to present our results with this constraint on the times. Nonetheless, our analysis shows that free-fall time of 100~s would be achievable within the parameters of the QPPF without spoiling the interference pattern.

\begin{figure*}[t!]
\centering
\includegraphics[width=1.5\columnwidth]{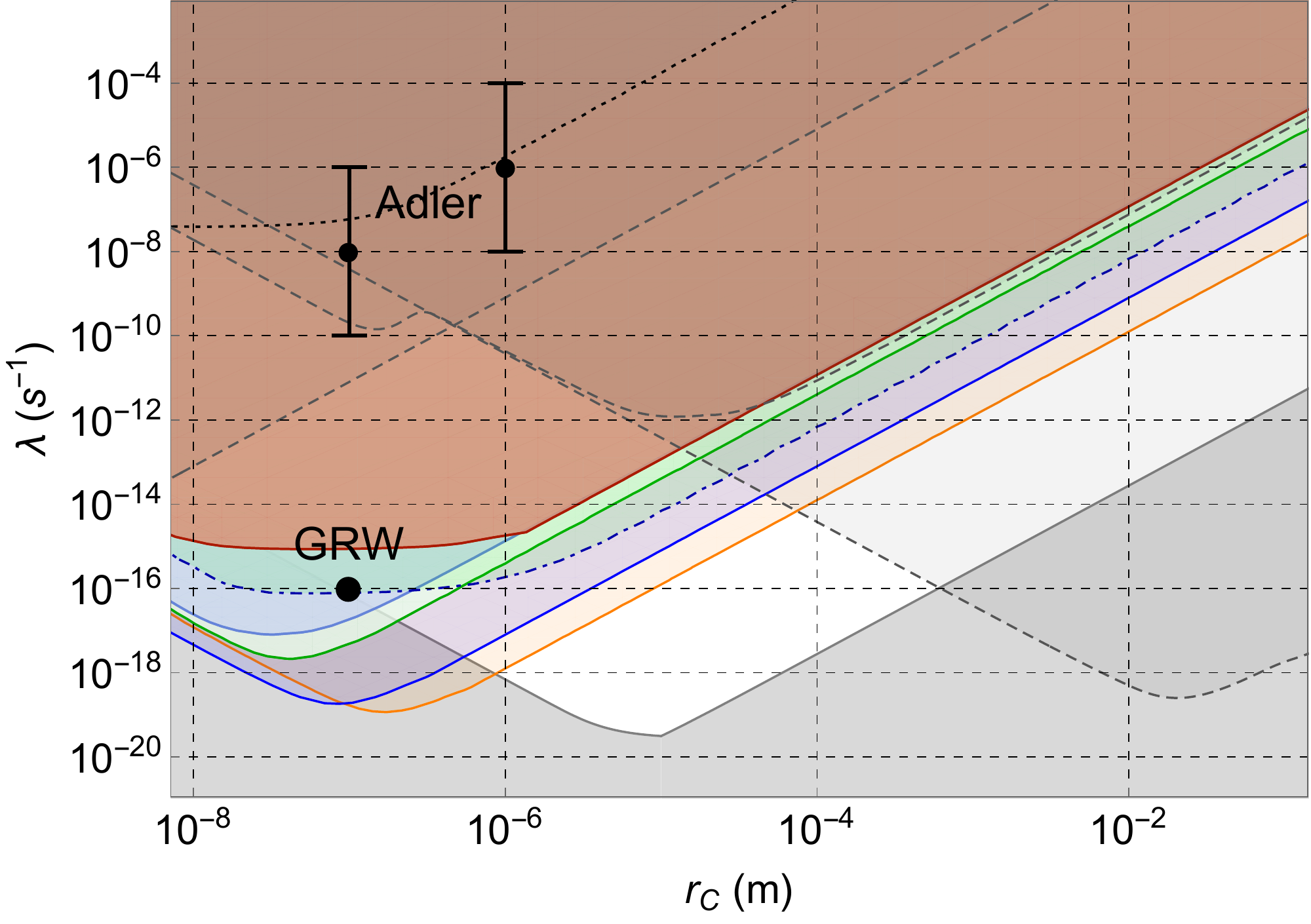}
\caption{Exclusion plots for the CSL parameters $\{r_c,\,\lambda_\text{CSL} \}$ {from interferometric experiments}. Black points, with their error bars, represent the GRW’s and Adler’s theoretical values for these parameters~\cite{ghirardi1986unified,adler2007collapse,adler2007corrigendum}.
The solid lines, and their respective excluded areas, show the upper bound that can be obtained by near field interferometric experiments in space with the parameters used in our simulations. In particular, the red line is obtained with $m=10^7 $~amu, $t_1=t_2=12$~s and $E_{L}/a_{L} = 1.1\times 10^{-2}$~$J/m^2$. The dark-green one with $m=10^8$~amu, $t_1=t_2=10$~s and $E_{L}/a_{L} = 3.5\times 10^{-4}$~J/m$^2$. The green, solid line with $m=10^9$ amu, $t_{1}=t_{2}=10$~s and $E_{L}/a_{L} = 8.7\times 10^{-6}$~J/m$^2$. The blue one with $m=10^{10}$~amu , $t_{1}=t_{2}=50$~s and $E_{L}/a_{L} = 8.7\times 10^{-6}$~J/m$^2$ and, finally, the orange solid line with $m=10^{11}$~amu , $t_{1}=t_{2}=50$~s and $E_{L}/a_{L} = 2.2\times 10^{-5}$~J/m$^2$. 
For comparison, the blue dot-dashed line shows the upper bound that could be achieved with a ground-based interferometric experiment with $m=10^7$~amu and times $t_1=t_2=9.95$~s -- free-falling times which are clearly beyond current reach for ground-based experiments -- as discussed in a recent study of four of the authors~\cite{gasbarri2020prospects}.
Finally, the dashed grey line represents the upper bounds given by current non-interferometric tests~\cite{vinante2020narrowing,piscicchia2017csl,carlesso2016experimental,vinante2016upper}, the black dotted lines on the top of the figure are the current upper bounds coming from interferometric tests~\cite{fein2019quantum,carlesso2019collapse,torovs2017colored,kovachy2015quantum}, and the grey area at the bottom of the plot represents the theoretical lower-bound provided in Ref.~\citeonline{torovs2018bounds}.}\label{fig:csl_excl}
\end{figure*}
{As outlined above, the first step in the analysis is to consider when $\aleph_{QC}$ is large enough to guarantee the possibility to certify a quantum mechanical interference pattern and then consider the corresponding $\aleph_{QCSL}$. Figure~\ref{fig:QMCSLYet} shows the results of our numerical investigation in this respect. The panels in the first row show the values of $\aleph_{QC}$, i.e., the distance between the classical shadow pattern and the quantum interference one, for particles masses $\{10^7,10^8,10^9,10^{10},10^{11}\}$~amu as a function of $t_1,t_2$ and for the values of $E_L/a_L$ which maximize the distinguishability. The latter are reported, as a function of $t_1,t_2$, in the SM~\cite{SI} (see Fig.~2 therein). From the first row of Fig.~\ref{fig:QMCSLYet} we see that $\aleph_{QC}$ takes values definitely larger than the experimentally justifiable threshold of $5\%$ for free-fall times  $t_1+t_2\leq 100$~s, opening the way to direct tests of the quantum superposition principle with mesoscopic quantum systems in large spatial superpositions. The panels in the second row in Figure~\ref{fig:QMCSLYet} show instead the comparison between the quantum interference pattern and the one which would arise if the CSL noise -- with parameters chosen at $\lambda_\text{CSL} =10^{-8}$~s$^{-1}$ and $r_c=10^{-7}$~m as proposed by Adler~\cite{adler2007lower} -- was present. The panels on the second row are obtained by  evaluating the cost function $\aleph_{QCSL}$ at the same values of $E_L/a_L$ used for the upper row, i.e. the values that, at fixed $\{t_1,t_2\}$, maximize the quantum-to-classical distinguishability. 
It should be noted that, for the comparison between CSL and quantum mechanics, we do not necessarily need to restrict our attention to only the values of the parameter $E_L/a_L$ that maximize the classical-quantum distinguishability. Indeed, by direct inspection of the interference figures it can be deduced that, in general, the classical and CSL patterns are quite different as far as they are not both flattened out by the effects of the noises (environmental or fundamental). This means that we can look for other parameters values which increase the distance between the quantum and CSL patterns. We show this on the third row of Fig.~\ref{fig:QMCSLYet} where we report the values of $\aleph_{QCSL}$ at the values of $E_L/a_L$ which maximize it. As it can be seen, the difference with the panels of the second row is not large apart for very light masses, meaning that the combined maximum distinguishability is nearly achievable.  
\\
Finally, Figure~\ref{fig:csl_excl} extends the previous analysis to the whole parameter space of the CSL model. This exclusion plot is obtained for values of the parameters $t_1,t_2,E_L/a_L$ which maximize the distinguishability between the quantum and CSL predictions, i.e., $\aleph_{QCSL}$, as shown in the third row of Figure~\ref{fig:QMCSLYet}. The solid lines in Figure~\ref{fig:csl_excl} show the upper bounds that could be achieved with space-based near-field interferometry experiments with particle masses up to $10^{11}$~amu. As it can be seen, already the use of  $10^9$~amu particles (green solid line) has the potential to rule out collapse models even beyond the values GRW originally proposed for the parameters, a feat that is outside the reach of current experiments. 
This is one of the main results of this work. It shows that near-field space-based experiments holds the promise to push tests of quantum mechanics -- and of collapse models -- way beyond what is possible with ground-based experiments and have the ability {to directly access a large and unexplored area of parameter space \{$\lambda_\text{CSL} ,r_{c}$\} of } the considered modifications of quantum  mechanics. 
}

In conclusion we should cite that, while the analysis presented in this section makes use of the formalism developed to account for the finite size of the particles~\cite{PhysRevA.100.033813} , and we have included all major sources of decoherence following the technical details laid down in the ESA's QPPF report~\cite{QPPF}, the description of the system suffers from an unavoidable level of idealization. Without entering in the discussion of technical challenges like the load and re-use of the nanoparticles in several runs of the experiment, we can still point out some of the idealizations made that enter directly into the simulations of the interferometric set-up. In particular, throughout this work we have assumed: the particles to be perfectly spherical, thus neglecting rotational degrees of freedom; the particles to be homogeneous, which has allowed us to use the formalism~\cite{PhysRevA.100.033813} derived from Mie-scattering theory; finally we have employed the sphere's bulk material refraction index which is tabulated in the literature. This last point is discussed in some detail in Ref.~[\citeonline{PhysRevA.100.033813}], where it is shown how the coherence properties of the grating interaction strongly depend on the refraction index. It is thus a crucial step for any realization of interferometric space-based experiments with large  nanoparticles to conduct preliminary experiments to determined the physical properties of the nanoparticles, with particular reference to their refractive index which could deviate from the bulk material one.


 \section{Conclusion and Outlook}\label{WishList}
In this perspective article, we have discussed the unique possibilities offered by the space environment for investigating the quantum superposition principle by dedicated interfrometric and non-interferometric experiments and to test quantum mechanics in the parameter regime of large-mass particles, impossible to reach on ground by today's technology. In particular, we have focused our attention on the generation and certification of spatial quantum superpositions of particles with sizes of the order of hundreds of nanometers and the possibilities that this offers for fundamental tests of quantum theory and alternatives thereof~\cite{millen2020quantum}.

After arguing for the advantages offered by space, being the long free-fall times and the availability of low-noise conditions, we considered two main experimental strategies for fundamental studies in space. The first one is the indirect approach of non-interferometric experiments, which does not require the creation of the spatial superposition. This strategy has been proven key in recent work on ground to test collapse models in otherwise unreachable parameter regimes. The second strategy is the more direct one based on interferometric experiments. Here, near-field interferometry with large dielectric nanospheres is the current powerhouse, proven experimental technology, and shows its potential when combined with the advantages of the space environment. We have reported a detailed forecast of the potential offered by these techniques based on state-of-the-art parameters values and showed how space-based experiments offer the possibility to both certify the creation of macroscopic superpositions and essentially rule-out an entire family of alternative models to standard quantum mechanics. Most importantly, we have not found a fundamental showstopper for performing both interferometric and non-interferometric experiments in space.

Needless to mention, large spatial superpositions of high-mass systems will provide a fine probe for further tests of fundamental physics. This includes: the domain of high-energy particle physics beyond the standard model, when it comes to test candidates of Dark Matter~\cite{riedel2013direct, bateman2015existence, riedel2017decoherence, carney2019ultralight, carney2020mechanical, monteiro2020search} and possible effects in particle interactions related to Dark Energy~\cite{khoury2004chameleon, rider2016search, moore2014search}; the low-energy regime of the interplay between quantum mechanics and gravity~\cite{pikovski2015universal, torovs2017quantum, belenchia2016testing, belenchia2019tests, carlesso2019testing}; precision tests of gravity~\cite{qvarfort2018gravimetry, hebestreit2018sensing, belenchia2018quantum, bose2017spin, hu2008stochastic, bahrami2014role, bassi2017gravitational, grossardt2016optomechanical}; the test of the equivalence principle and of general relativity's predictions, such as gravitational waves, in a parameter range complementary to existing experiments such as LIGO, VIRGO, GEO600 and the planned LISA space antenna~\cite{arvanitaki2013detecting, pontin2018levitated}, and frame-dragging effects~\cite{fadeev2020gravity}. Furthermore, large-mass experiments in space will unavoidably provide a formidable platform for applications in Earth and planet observation~\cite{middlemiss2016measurement, banerdt2020initial}, where large-mass mechanical systems have already shown a superb capability as force and acceleration sensors~\cite{millen2020optomechanics,li2018characterization, fadeev2020ferromagnetic, vinante2020ultralow, mitchell2020colloquium, timberlake2019static, hempston2017force, ranjit2016zeptonewton, ahn2020ultrasensitive, geraci2010short}, including in rotational mechanical modes~\cite{carlesso2018non, stickler2018probing}; as well as their use for frequency conversion~\cite{forsch2020microwave}\footnote{\MP{Whilst in a different set-ups with respect to the ones considered here.}} 

It is clear that the realisation of large-mass, fundamental physics experiments in space is an immensely challenging project. Therefore, the most important next step is to form a community of scientists, industry and space agencies for defining a concrete road-map for the accomplishment of a successful space mission by working on fine-tuned theoretical analysis of conditions for the experiment, coming up with new proposals to test further new physics in the large-mass regime and, last but not least, to push the development of technology readiness for space. Such a roadmap should include performing proof-of-principle large-mass experiments in micro-gravity environments -- such a sounding rockets, drop towers and Einstein elevators, space stations, cubesats, and potentially on the Moon and Mars -- in alignment with international and national space agencies plans for future fundamental physics experiments in space. We hope that the results of this work will stimulate the physics community to further investigate the possibilities offered by space-based experiments of this kind.

\section*{Acknowledgements}
A. Belenchia acknowledges support from the MSCA IF  project pERFEcTO  (Grant  No. 795782) and the Deutsche Forschungsgemeinschaft (DFG, German Research Foundation) project number BR 5221/4-1. A. Bassi acknowledges financial support from the INFN, the University of Trieste and the support by grant number (FQXi-RFP-CPW-2002) from the Foundational Questions Institute and Fetzer Franklin Fund, a donor advised fund of Silicon Valley Community Foundation.
S. Donadi acknowledges financial support from INFN and the Fetzer Franklin Fund. G. Gasbarri acknowledges support from the Spanish Agencia Estatal de Investigaci\'{o}n, project PID2019-107609GB-I00, from the QuantERA grant C'MON-QSENS!, by Spanish MICINN PCI2019-111869-2, and from COST Action CA15220. {R. Kaltenbaek} acknowledges support by the Austrian Research Promotion Agency (projects 854036, 865996) and by the Slovenian Research Agency (research projects N1-0180, J2-2514, J1-9145 and P1-0125).  M. Paternostro is supported by the DfE-SFI Investigator Programme (grant 15/IA/2864), the Royal Society Wolfson Research Fellowship (RSWF\textbackslash R3\textbackslash183013), the Leverhulme Trust Research Project Grant (grant nr.~RGP-2018-266), and the UK EPSRC (grant nr.~EP/T028106/1). H. Ulbricht acknowledges support from The Leverhulme Trust (RPG-2016- 046) and the UK EPSRC (EP/V000624/1). A. Bassi, M. Carlesso, M. Paternostro, and H. Ulbricht are supported by the H2020 FET Project TEQ (Grant No. 766900). All the authors acknowledge partial support from COST Action QTSpace (CA15220) and thank  Alexander Franzen for the creation of one of the figures in the manuscript.

\section*{Author contributions}
G. Gasbarri and A. Belenchia led the development of the project with strong input from M. Carlesso and S. Donadi and with significant contributions from A. Bassi, R. Kaltenbaek, M. Paternostro, and H. Ulbricht. All authors contributed to the preparation of the manuscript and its finalisation.

\section*{Competing Interests}
The authors declare no competing interests.

\bibliography{sample}
\end{document}


\maketitle

\setcounter{equation}{0}
\setcounter{figure}{0}
\setcounter{table}{0}
\setcounter{page}{1}
\makeatletter
\renewcommand{\theequation}{S\arabic{equation}}
\renewcommand{\thefigure}{S\arabic{figure}}

\section{Interferometric Experiments: Theory}

\subsection{Generalized Talbot Coefficients}
We report here the explicit expressions of the generalized Talbot coefficients. We refer the interested reader to Ref.~[\citeonline{PhysRevA.100.033813}] and reference therein for the detailed derivation of these expressions and the analysis of the Talbot effect. 

The generalized Talbot coefficients encode encode the coherent and incoherent effects of the optical grating on the state of the nanoparticle. Assuming the waist of the standing wave in the grating to be much bigger than the matter-wave profile, and for short interaction times ($\tau_{int}$) compared to the characteristic time of spreading of the matter-waves, the evolution of the nanoparticle density matrix in the longitudinal position representation is given by
\begin{align}\label{eq:trgrat}
\bra{z}\rho\ket{z'}\rightarrow R(z,z')T(z,z')\bra{z}\rho\ket{z'},
\end{align}
where  
\begin{align}\label{Tco}
&T(z,z')=t(z)t^{*}(z')=e^{-\frac{i}{\hslash}\int_{0}^{t}d\tau [V(z,\tau)-V(z',\tau)]}\\
&R\left(z,z'\right)= e^{\int_{0}^{\tau_{int}} d\tau \mathcal{L}_{t}(z,z')}=R_{\rm{sca}}(z,z')R_{\rm{abs}}(z,z').
\end{align}
with $V(z,t)$ the classical interaction potential. The term $T(z,z')$ describes the coherent effect of the grating while $R(z,z')$ the incoherent combined effect of scattering and absorption of grating photons.
The generalized Talbot coefficients entering the interference pattern are given by the convolution between the Talbot coefficient describing the sole coherent grating $B^{\rm{coh}}_{n}\left({s}/{d}\right)= \sum_{k}b_{k}b^{*}_{k-n}\exp\{{i \pi (n-2k)s}/d\}$, which are given in terms of 
the Fourier coefficients $b_{n}$ of the transmission function $t(z)$ entering the coherent transmission mask $T(z,z')$, and the Fourier coefficients $R_{n}$ of the decoherence mask function $R(z,z')$  
\begin{align}\label{eq:TalbotCoeff}
{B}_{n}\left(\frac{s}{d}\right)&=\sum_{j}B^{\rm{coh}}_{n-j}\left(\frac{s}{d}\right)R_{j}\left(\frac{s}{d}\right)\\
&= e^{F-c_{\rm{abs}}/2} \sum_{k=-\infty}^{\infty}\left(\frac{\zeta_{\rm{coh}}+a+c_{\rm{abs}}/2}{\zeta_{\rm{coh}}-a-c_{\rm{abs}}/2}\right)^{\frac{n+k}{2}}J_{k}(b)J_{n+k}\left(\sign(\zeta_{\rm{coh}}-a-c_{\rm{abs}}/2)\sqrt{\zeta_{\rm{coh}}^{2}-(a+c_{\rm{abs}}/2)^{2}}\right).
\end{align}
Here    
\begin{align}\label{abF}
c_{\rm{abs}}&=
\frac{4\sigma_{abs}}{ h c}\frac{E_{L}}{a_{L}}(1-\cos(\pi s/d)\nonumber\\
\zeta_{coh} &= \frac{16F_{z}(-\lambda/8) E_L}{\hslash a_L k I_{0}}\sin\left(\frac{\pi s}{d}\right)\nonumber\\
a(s) &= \int d\tau \frac{8\pi  E_{L}}{\hslash \omega a_{L}}\int d \Omega\,{\rm Re}\big(f^{*}(k , k \mathbf{n})f(- k , k\mathbf{n})\big)[\cos(k n_{z} s)-\cos(ks)],\nonumber\\
b(s)&=  \int d\tau\frac{8\pi  E_{L}}{\hslash  \omega a_{L}}\int d \Omega \,{\rm Im} \big(f^{*}(k , k \mathbf{n})f(- k , k\mathbf{n})\big)\sin(kn_{z}s),\nonumber\\
F(s)&=  \int d\tau \frac{8 \pi E_{L}}{\hslash  \omega a_{L}}\int d \Omega\, |f(k,k\mathbf{n})|^{2} [\cos((1-n_{z})ks)-1],
\end{align}
where $\lambda= 2\pi / k= 2\pi /\omega c $ is the light wavelength, $E_{L}$ is grating laser energy, $I_{0}$ and $a_{L}$ are respectively the intensity parameter and the spot area of the laser, $F_{z}(z)$ is the longitudinal light-induced force on the dielectric sphere, $\sigma_{abs}$ is the photon absorption cross section, and $f(k,k\mathbf{n})$ the photon scattering cross section.

\subsubsection{Classical Limit of the optical grating}\label{ClaLim}
As we discussed in the main text, also classical particles following ballistic trajectories through the grating laser would form a near-field shadow pattern. The effect of the optical grating on classical particles can be described by modifying in a suitable way the generalized Talbot coefficients. In particular, the classical behaviour can be obtained as the limiting case of $\hslash \to 0$ as outlined in Ref.~[\citeonline{PhysRevA.100.033813}]. To summarize, the classical analogue of the coherent effect of the grating corresponds to a momentum kick due to the classical optical force acting on a dielectric and, by performing an expansion of the trigonometric functions in $a,b, F$ to first order in $\hslash$, it is easy to see that the only non-vanishing contribution to the classical coefficients arise from the function $b$. 

\subsection{Environmental Decoherence}
As discussed in the main text, our simulations account for several sources of environmental decoherence acting during the free fall times in the experiment.  In particular, we account for decoherence due to collision with the residual gas in the vacuum chamber; black-body radiation, emission, scattering, and absorption -- taking into account the heating of the particle (photonic environment) during the trapping period and the subsequent cooling during free-fall. Here we report the expression for the kernel $R_{n}$ in Eq.~(3) in the main text that account for all these decoherence sources.
\begin{align}
\ln(R_{n})=& -\Gamma_{\text{coll}}(t_{1}+t_{2})+\int d\omega \gamma_{\text{abs}}(\omega)\left[\frac{\text{Si}(a_{n})}{a_{n}}-1\right](t_{1}+t_{2})+\int d\omega \gamma_{\text{sca}}(\omega)\left[\frac{\text{Si}(2a_{n})}{a_{n}}-\text{sinc}^{2}(a_{n})-1\right](t_{1}-t_{2})\nonumber\\
&\int d\omega \int_{0}^{1} d\theta \{t_{1}\gamma_{\text{emi}}[\omega,T_{\text{int}}(t_{1}-t_{1}\theta)]+t_{2}\gamma_{\text{emi}}[\omega,T_{\text{int}(t_{1}+t_{2}\theta)] }[\text{sinc}(a_{n}\theta)-1]
\end{align}
with $a_{n}= n\,h\,\omega\, t_{2} t_{1} / ((t_2+t_1)m\,c\,d) $, and $\text{Si}$ the sine integral. The collision rate $\Gamma_\text{coll}$ is given by
\begin{align}
\Gamma_{\text{coll}} = \frac{ 4\pi \Gamma(9/10)}{5\sin(\pi/5)}\left(\frac{3\pi C_{6}}{2\hslash}^{2/5} \frac{p_{g}v_{g}}{k_{\text{B}}T_{\text{env}}}\right)
\end{align}
while the scattering, emission and absorption rates by
\begin{align}
\gamma_{\text{sca/abs}}(\omega) =\frac{ (\omega / \pi c)^{2}\sigma_{\text{sca/abs}}(\omega)}{\exp(\hslash \omega/k_{\text{B}}T_{\text{env}})-1},\hspace{1cm}
&\gamma_{\text{emi}}[\omega, T_{\text{int}}]= \left(\frac{\omega}{\pi c}\right)^{2}\sigma_{\text{abs}}\exp\left(-\frac{\hslash \omega}{k_{\text{B}}T_{\text{int}}}\right)\text{Im}\left\{\frac{\varepsilon(\omega)-1}{\varepsilon(\omega)+2}\right\}
\end{align}
where  $v_{g}$ and $p_{g}$ are the mean velocity and pressure of the gas, $T_{\text{env}}$  the environmental temperature, $\sigma_{\text{abs/sca}}$ the photon scattering/absorption cross section, $\varepsilon(\omega)$ the electric permittivity, and 
\begin{align}
C_{6} \simeq \frac{ 3 \alpha(\omega=0) \alpha_{g} I_{g}I}{32 \pi^{2}\varepsilon_{0}^{2}(I_{g}+I)}
\end{align}
the van der Waals coupling constant where $\alpha$, $\alpha_{g}$ are the static polarizabilities and $I$, $I_g$ the ionization energies of the nanosphere and the gas particle, respectively.

We refer the reader to the supplemental information of~[\citeonline{bateman2014near}] and the reference therein for a detailed derivation of these expressions.

\subsection{ CSL-Decoherence}
The dissipative term describing the effective decoherence of the center-of-mass wavefunction -- for the reduced one-dimensional state of motion of a nanoparticle -- due to the CSL reads
\begin{align}\label{cslme}
\mathcal{L}_{CSL}(x,x')=-\frac{\lambda_{CSL}(4\pi r_C^2)^{3/2}}{(2\pi\hslash)^{3}}\int d\mathbf{q}\,\frac{\tilde{\mu}(\mathbf{q})^2}{m_0^2}\,e^{-r_C^2\mathbf{q}^{2}/\hslash^{2}}\,\left(e^{-\frac{i}{\hslash}{q}_{z}({x}-{x'})}-1\right)
\end{align}   
in position representation. Here $m_{0}$ the nucleon mass , $\lambda_{CSL}$ and $r_{c}$ the rate and the localization length of the CSL model, and $\mu(\bf{q})$ the Fourier transform of the nano-particle mass density $\mu(\bf{x})$
,i.e.
\begin{align}
\tilde{\mu}(\mathbf{q}):= \int e^{-\frac{i}{\hslash} \mathbf{q}\cdot\mathbf{x}}\mu(\mathbf{x}).
\end{align}
that in the case of a homogeneous and spherical mass distribution of radius $R$ is given by
\begin{align}
 \tilde{\mu}(\mathbf{q})= \frac{4 \pi \hslash R^2}{q} J_{1}(q R/\hslash)
\end{align}
where $J_{1}(q)$ denotes the spherical Bessel function of the first kind.  
Exploiting this equation, rewriting the integral in Eq.~\eqref{cslme} in spherical coordinates, and performing the integration over the solid angle the CSL dissipative term simplifies to 
\begin{align}
\mathcal{L}_{CSL}(\mathbf{x},\mathbf{x}')= \frac{64 \pi^{3/2}\,\lambda_{CSL}\, rc^{3} R^{4}}{\hslash\, m_{0}^{2}}\int_{0}^{\infty} d q\, e^{-r_{c}^2 q^2/\hslash^2} J_{1}(q R/\hslash)^{2}\left(\text{sinc}(q|{x}'-{x}|)/\hslash)-1 \right).
\end{align}
This results in the following, additional kernels entering the expression for the interference pattern probability:
\begin{align}
R_{n}^{\text{CSL}}= \exp\left\{ \Gamma_{\text{CSL}}\left(f_{\text{CSL}}\left(\frac{ h\, n\, q\, t_1 t_2}{m\, d (t_1+t_2)}\right)-1\right)(t_1+t_2)\right\}
\end{align}
where $d$ is again the grating period and 
\begin{align}
\Gamma_{\text{CSL}}&=  \sqrt{\frac{32}{\pi}}\frac{\lambda_{\text{CSL}}\, r_{c}^{3}}{\hslash^{3} m_{0}^{2}} \int dq q^2 e^{-r_{c}^{2}q^2/\hslash^2}\tilde{\rho}(q)^{2}  \nonumber\\
f_{\text{CSL}}(x)&=\sqrt{\frac{32 }{\pi}} \frac{\lambda_{CSL}\,r_{c}^{3}}{\hslash^3 m_{0}^{2} \Gamma_{\text{CSL}}}\int dq q^{2} e^{-r_{c}^{2}q^{2}/\hslash^{2}} \tilde{\rho}(q)^{2} \text{Si}\left(\frac{q x}{\hslash}\right) =\frac{64 \pi^{3/2}\lambda_{CSL} rc^{3} R^{4} }{\hslash m_{0}^{2} \Gamma_{\text{CSL}}} \int_{0}^{\infty}dq e^{-q^2 r_{c}^{2}/\hslash^{2}}J_{1}\left(\frac{q R}{\hslash}\right)^{2}\text{Si}\left(\frac{qx}{\hslash}\right).
\end{align}
\begin{figure*}
\centering
\includegraphics[width=1.\textwidth]{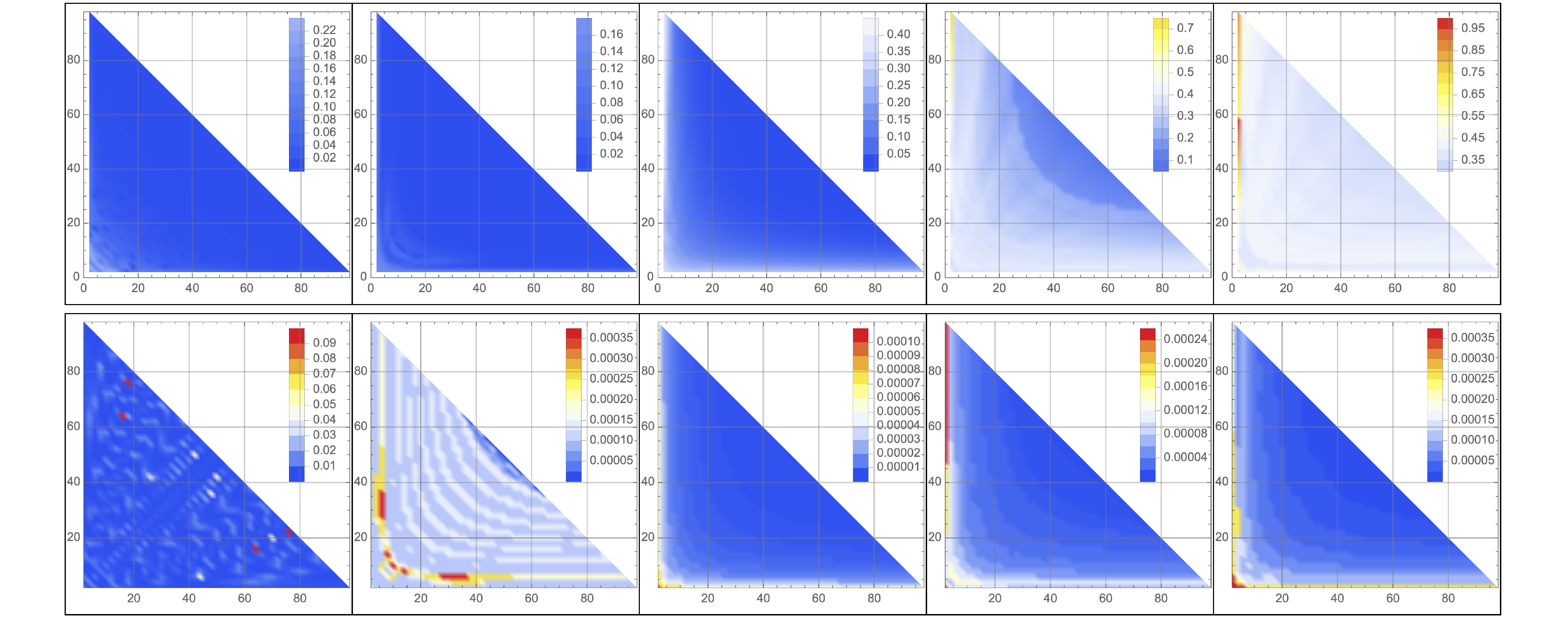}
\caption{Comparison between quantum mechanics prediction and CSL one. In particular, the first row of the panel shows the cost function $\aleph_{QCSL}$, for increasing mass from left to right, as a function of $t_1, t_2$. The second row shows the values of the $E_{L}/a_{L}$ ratio that maximise the cost function $\aleph_{QCSL}$. 
}\label{fig:csl_cost_del}
\end{figure*}

\begin{figure*}
\centering
\includegraphics[width=1.\textwidth]{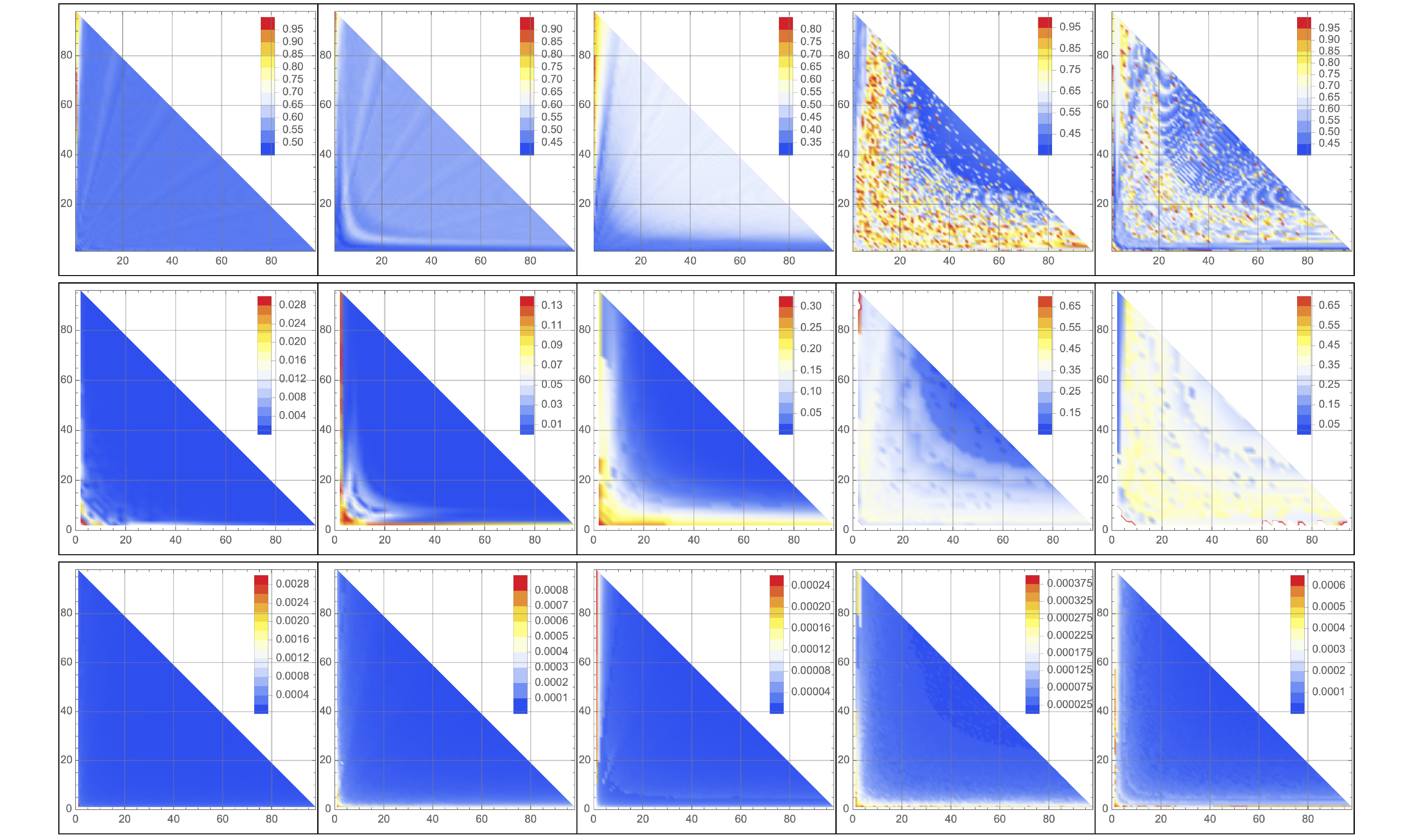}
\caption{Comparison between quantum mechanics, classical shadow pattern and the prediction of CSL. The first row shows the value of $\aleph_{QC}$ for increasing value of the mass while the second row shows the value of $\aleph_{QCSL}$. The third row shows instead the values of the $E_{L}/a_{L}$ that have been used in both the first and second row figures. These values of the $E_{L}/a_{L}$ ratio are the one which maximize $\aleph_{QC}$. 
}\label{fig:QMCSL}
\end{figure*}


\section{Interferometric Experiments: Simulations}
Here we report additional details for the simulations performed for near-field intereferometric experiments in space. In Table~\ref{table} we report all the relevant parameters used in the simulations leading to the results in section~IV.

It should be noted that, in the results shown, we did account for the heating of the nanoparticle while initially optically trapped for a time $t_c$ by assuming an internal temperature of $40$~K after time $t_c$ in an environment at $20$~K. This is a conservative estimate and we also run the same simulations for higher and lower values of the initial internal temperature without noticing any significant deviation. 

\begin{table}[h!]
\centering
\begin{tabular}{llll}
    \hline
    \textbf{Symbol}  & & \textbf{Name} & \textbf{Value}  \\ 
    &&&{\bf or Expression}\\
    \hline\hline
    Nanosphere\\ properties:\\ \hline
    $\rho_{SiO2}$ & & Glass SiO2 density & $1850$~kg/m$^3$ \\
    $c_{m}$ & & Specific heat  & 700~J/(kg K) \\
    $I$ & & Ionization energy & $5\times 10^{-19}$~J \\
    $m$ & & Mass & $10^7 - 10^{11}$~amu \\ 
    $\sigma_x$ & & Position variance & $\sqrt{\frac{\hslash}{4\gamma}\coth{({\beta_0\nu_m})}}$ \\
    &&post-cooling &\\
    $\sigma_p$ & & Momentum variance& $\sqrt{\hslash\gamma\coth{({\beta_0\nu_m})}}$ \\
    &&post-cooling &\\
    \hline 
    Trapping\\ parameters:  \\
    \hline
    $\lambda_c$ & & Trap's laser wavelength & 1550~nm \\
    $t_c$ & & Trapping and cooling time & 1~s \\
    $I_{{\rm trap}}$ & & Trap's laser intensity & $90\times 10^9$~W/m$^2$ \\
    $\nu_{m}$ & & Trap mechanical frequency  &  $10^5/2\pi$~Hz \\
     &&--longitudinal direction &\\
    $T_{{\rm int}0}$ & & C.o.m. temperature  &  $\sim 5\times 10^{-6}$~K\\
    &&post-cooling &\\
    \hline
    Optical grating\\ parameters: \\ \hline
    $\lambda$ & & Grating laser wavelength & $100$~nm \\
    $E_L/a_L$ & & Pulse-energy per spot-area & $10^{-6}- 5$ J/m$^{2}$  \\ 
    \hline
    Environment\\ parameters: \\ \hline
    $T_{\rm env}$ & & Environmental temperature & 20~K \\
    $\alpha_g/(4\pi\epsilon_0)$ & &  Residual gas polarizability & $0.6668\times 10^{-30}$~m$^3$ \\
    $I_g$ & & Residual gas ioniz. energy & $2.17\times 10^{-18}$~J\\
    $m_g$ & & Residual gas mass & 1.00784~amu\\
    $v_g$ & & Residual gas mean velocity  & $\sqrt{2k_B T_{\rm env}/m_g}$\\
    $P_g$ & & Residual gas pressure & $10^{-13}$~mbar \\ \hline
\end{tabular}
\caption{Specifics of all the parameters entering the simulation of the near-field interferometric experiments with dielectric nanospheres. The residual gas in the vacuum chamber is assumed to be composed mostly by nitrogen, of which we use the physical properties. The refractive index, as a function of the frequency, for Si and SiO2 can be found tabulated in the supplementary material of Bateman et al.~\cite{bateman2014near} (see also references therein). The pulse energy to spot area ratio of the grating laser is related to the eikonal phase by Eqs.~(20-22) in Ref.~[\citeonline{PhysRevA.100.033813}] to which we refer the reader for additional details. 
We have set $\beta_0=h/(2k_B T_{{\rm int}0})$ and $\gamma=\pi m\nu_m$.}\label{table}
\end{table}

In Figure~\ref{fig:csl_cost_del} we report the values of $\aleph_{QCSL}$ (first row) for different masses of the nanoparticles and as a function of $t_1, t_2$. This correspond to the panels in the third row of Figure~6 in the main text. The panels on the second row of the figure show the values of $E_L/a_L$ used. These are the values that maximize $\aleph_{QCSL}$.

In Figure~\ref{fig:QMCSL} we report instead both the values of $\aleph_{QC}$ (first row) and the ones of $\aleph_{QCSL}$ (second row) obtained using the values of $E_L/a_L$ shown in the third row which that maximize $\aleph_{QC}$.

\newpage
\bibliography{sample}